\documentclass[aps,prb, twocolumn,showpacs,superscriptaddress]{revtex4-1}
\usepackage{dsfont,amsthm,amsbsy}
\usepackage{amssymb}
\usepackage{amsmath}
\usepackage{bbm}
\usepackage{graphicx}
\usepackage{epstopdf}
\usepackage{subfigure}
\usepackage{natbib}
\usepackage{epsfig}
\usepackage{amsfonts}
\usepackage{mathrsfs}
\usepackage{sidecap}
\usepackage{lipsum}
\usepackage{xcolor}
\usepackage[toc,page,title,titletoc,header]{appendix}
\usepackage[colorlinks,linkcolor=blue,citecolor=blue,anchorcolor=blue, urlcolor=blue]{hyperref}
\usepackage{hyperref}
\usepackage{resizegather}
\usepackage{float}
\usepackage{mathbbol}
\usepackage[normalem]{ulem}
\usepackage{cancel}
\usepackage{upgreek}
\usepackage{graphics,dcolumn,bm}
\usepackage{rotate,color}
\usepackage{times}
\newcommand\redsout{\bgroup\markoverwith{\textcolor{red}{\rule[0.5ex]{2pt}{0.4pt}}}\ULon}

\newcommand{\bl}{\begin{aligned}}
\newcommand{\el}{\end{aligned}}
\def\be{\begin{equation}}
\def\ee{\end{equation}}

\def\bi{\begin{itemize}}
\def\ei{\end{itemize}}
\def\bn{\begin{enumerate}}
\def\en{\end{enumerate}}
\def\bea{\begin{eqnarray}}
\def\eea{\end{eqnarray}}
\def\no{\nonumber}
\def\ba{\begin{array}}
\def\ea{\end{array}}
\def\bd{\begin{displaymath}}
\def\ed{\end{displaymath}}

\begin{document}

\title{Floquet dynamical quantum phase transition in the extended XY model: \\
nonadiabatic to adiabatic topological transition}
\author{Sara Zamani}
\email{sara.zamani@iasbs.ac.ir}
\affiliation{Department of Physics, Institute for Advanced Studies in Basic Sciences (IASBS), Zanjan 45137-66731, Iran}
\author{R. Jafari}
\email[]{jafari@iasbs.ac.ir, rohollah.jafari@gmail.com}
\affiliation{Department of Physics, Institute for Advanced Studies in Basic Sciences (IASBS), Zanjan 45137-66731, Iran}
\affiliation{Department of Physics, University of Gothenburg, SE 412 96 Gothenburg, Sweden}
\affiliation{Beijing Computational Science Research Center, Beijing 100094, China}
\author{A. Langari}
\email[]{langari@sharif.edu}
\affiliation{Department of Physics, Sharif University of Technology, P.O.Box 11155-9161, Tehran, Iran}

\begin{abstract}
We investigate both pure and mixed states Floquet dynamical quantum phase transition (DQPT) in the periodically time-dependent extended XY model. We exactly show that the proposed Floquet Hamiltonian of interacting spins can be expressed as a sum of noninteracting quasi-spins imposed by an effective time dependent magnetic field (Schwinger-Rabi model).
The calculated Chern number indicates that there is a topological transition from nonadiabatic to adiabatic regime. In the adiabatic regime, the quasi-spins trace the time dependent effective magnetic field and then oscillate between spin up and down states. While in the nonadiabatic regime, the quasi-spins cannot follow the time dependent effective magnetic field and feel an average magnetic field. We find the range of driving frequency over which the quasi-spins experience adiabatic cyclic processes. Moreover, we obtain the exact expression of the Loschmidt amplitude and generalized Loschmidt amplitude of the proposed Floquet system. The results represent that both pure and mixed states dynamical phase transition occurs when the system evolves adiabatically. In other words, the minimum required driving frequency for the appearance of Floquet DQPT is equal to the threshold frequency needed for transition from nonadiabatic to adiabatic regime.
\end{abstract}


\maketitle

\section{Introduction\label{introduction}}

Recent experimental advances on synthesizing various quantum platforms, including ultra-cold atoms in optical lattices \cite{jotzu2014experimental,daley2012measuring,schreiber2015observation,choi2016exploring,flaschner2018observation}, trapped ions \cite{jurcevic2017direct,martinez2016real,neyenhuis2017observation,smith2016many}, nitrogen-vacancy centers in diamond \cite{yang2019floquet}, superconducting qubit systems \cite{guo2019observation} and quantum walks in photonic systems \cite{wang2019simulating,xu2020measuring} provide a framework for experimentally studying different quantum systems far from thermodynamic equilibrium. Nonequilibrium physics with exotic properties, specifically  realizing the quantum time evolution beyond the thermodynamic equilibrium description as well as the dynamics of out-of-equilibrium quantum many-body criticality, has recently attracted attention of theoretical \cite{Dutta:2017717,Essler2016,Fogarty2020,Campbell2016,Polkovnikov2011,Mitra2018,jafari2019dynamical,bhattacharya2017emergent,halimeh2019dynamical,
vzunkovivc2018dynamical,heyl2013dynamical,budich2016dynamical,heyl2017quenching,weidinger2017dynamical} and experimental \cite{xu2020measuring,yang2019floquet,jurcevic2017direct,guo2019observation,wang2019simulating,xu2020measuring,smith2016many,martinez2016real,neyenhuis2017observation,schreiber2015observation,aidelsburger2011experimental,aidelsburger2013realization} researches in physics.

Lately, a new area of research named dynamical quantum phase transitions (DQPTs) \cite{heyl2013dynamical,heyl2018dynamical}, has been introduced, in non-equilibrium quantum systems, as a counterpart of conventional equilibrium thermal phase transitions. Within DQPT real time plays the role of control parameter analogous to temperature in conventional equilibrium phase transitions \cite{heyl2018dynamical,zvyagin2016dynamical}. DQPT represents a phase transitions between dynamically emerging quantum phases, that occurs during the nonequilibrium coherent quantum time evolution under quenching or time-periodic modulation of Hamiltonian \cite{heyl2018dynamical,zvyagin2016dynamical,heyl2018dynamical,yang2019floquet}.

The concept of DQPT originates from the analogy between the equilibrium partition function of a system, and boundary partition function, which measures the overlap between an initial state and its time-evolved one, termed as Loschmidt amplitude (LA).
As the equilibrium phase transition, characterized by nonanalyticities in the thermal free energy, in DQPT, nonanalytic behavior manifests as singularities in the LA, a dynamical analog of the equilibrium free energy \cite{heyl2018dynamical,zvyagin2016dynamical}.
Furthermore, nonanalyticities in dynamical free energy are accompanied by zeros of LA in the complex time plane known in statistical physics as Fisher zeros of the partition function \cite{heyl2013dynamical,heyl2018dynamical,vzunkovivc2018dynamical,jafari2019dynamical,halimeh2019dynamical,bhattacharya2017emergent,fisher1967theory,lee1952statistical,van1984location,zvyagin2016dynamical}.

It has been also established that there exists a dynamical topological order parameter (DTOP), analogous to order parameters at conventional quantum phase transition, which can characterize DQPTs \cite{budich2016dynamical,vajna2015topological,sedlmayr2018fate}. The presence of a DTOP illustrates the emergence of a topological characteristic associated with the time evolution of nonequilibrium systems. The DTOP takes integer values as a function of time and represent unit magnitude jumps at the critical times, signaling the occurrence of DQPTs \cite{sedlmayr2018fate,yang2019floquet,vajna2015topological,flaschner2018observation}.
Decoherence and particle loss processes do affect the dynamics, hence, the notion of DQPTs has been developed to the generalized form, i.e., thermal (mixed) state dynamical phase transition \cite{heyl2017dynamical,bhattacharya2017mixed,sedlmayr2018fate}.

DQPT has been extensively explored from both theoretical
\cite{Karrasch2013,halimeh2019dynamical,heyl2013dynamical,budich2016dynamical,
vzunkovivc2018dynamical,jafari2019dynamical,andraschko2014dynamical,sharma2015quenches,jafari2019quench,zhou2018dynamical,jafari2019dynamical,
andraschko2014dynamical,sharma2015quenches,canovi2014first,budich2016dynamical,bhattacharya2017emergent,
hickey2014dynamical,schmitt2015dynamical,Sun2020,Zhou2019,Mera2018,Khatun2019,
Srivastav2019,Abdi2019,Cao2020,Bhattacharyya2020,Ding2020,Rylands2020,Hu2020,Pastori2020,Kyaw2020}
and experimental \cite{flaschner2018observation,jurcevic2017direct,guo2019observation,wang2019simulating,yang2019floquet,martinez2016real,
lanyon2011universal,buyskikh2016entanglement,bernien2017probing,atala2014observation} point of views.
Most researches on DQPT, are associated with both sudden and slow quantum quenches of the Hamiltonian \cite{Sharma2016b,Sedlmayr2018,Sedlmayr2020,jafari2019dynamical,halimeh2019dynamical,bhattacharya2017emergent,budich2016dynamical,heyl2013dynamical,heyl2017quenching,vzunkovivc2018dynamical}. It was first found that, during the quench procedure, crossing the equilibrium quantum critical point (EQCP) is an essential condition to observe DQPT \cite{heyl2013dynamical}. However, subsequent analytical studies indicate that DQPT occurs regardless of any quenches across EQCP \cite{jafari2019dynamical,halimeh2019dynamical,vajna2014disentangling,sharma2015quenches}.

In addition, the theory of DQPT has been recently extended to Floquet systems \cite{kosior2018dynamical,kosior2018dynamical1,yang2019floquet}. In Floquet DQPT, as opposed to conventional quantum quench scenario, systems evolve under the time-periodic Hamiltonian \cite{yang2019floquet,kosior2018dynamical}. Although the quench-induced dynamical free energy accompanied with periodicity and decaying in time, in Floquet systems DQPT occurs periodically without decaying in time \cite{kosior2018dynamical,kosior2018dynamical1}.
Despite numerous studies of DQPTs in a wide variety of quantum systems, comparatively little attention has been devoted to the quantum Floquet systems. To the best of our knowledge, there is currently no conclusive analytical evidence that under which circumstances DQPTs appear in quantum Floquet systems. Further studies are needed to clarify and shed more light on this issue. For instance, exactly solvable models play a particularly important role in this direction.

In the present work, we contribute to expand the systematic understanding of Floquet DQPT by
introducing the exactly solvable periodically time-dependent extended XY model in the presence of a staggered magnetic field. We analytically study the pure and mixed states DQPT characteristics, especially the underlying topological features. We show that, pure state DQPT occurs when the system enters into the adiabatic regime. In other words, the minimum required driving frequency for appearance of Floquet DQPT is equal to the threshold frequency needed for transition from nonadiabatic to adiabatic regime. In the nonadiabatic regime, the quasi-spins (i.e. noninteracting effective spins) feel a constant effective Zeeman field, which does not induce Rabi oscillations between spin up and down states\cite{Schwinger1937}. When the system enters the adiabatic regime the quasi-spins oscillate between spin up and down states \cite{Schwinger1937}. We also indicate that the pure state dynamical free energy undergoes periodic nonanalyticities without decaying in time.  In contrast, the nonanalyticities of mixed state dynamical free energy are periodically robust in the presence of temperature,
where their amplitude decreases with time.
This implies mixed state Floquet DQPT time scale becomes the same as its corresponding pure state, which does not depend on the temperature. Moreover, discrete jumps of DTOP of pure and mixed states confirms consistently the topological feature of DQPT.
Although its topological nature is lost at finite temperatures.
It should be mentioned that, analogous to the message of
Ref. [\onlinecite{heyl2013dynamical}], the existence of Floquet DQPTs in Ref.[\onlinecite{yang2019floquet}]
is connected to the well-known equilibrium quantities given by two gapless critical points.
In contrast, the underlying model here, has a single critical point and the Floquet DQPTs
within a window frequency is shown to be related to the well-known dynamical notion i.e., adiabatic-nonadiabatic processes.

\section{Periodically Driven Extended XY Model\label{model}}

The Hamiltonian of the one-dimensional harmonically driven extended XY spin chain in the staggered magnetic field is given by
%
\begin{eqnarray}
{\cal H}(t) = \sum_{n=1}^{N}&&  J_{1} \Big[\cos(\omega t) \Big(S_n^x S_{n+1}^x + S_n^y S_{n+1}^y \Big)
\nonumber\\
&&- (-1)^{n} \sin(\omega t) \Big(S_n^x S_{n+1}^y - S_n^y S_{n+1}^x \Big)
\nonumber\\
&&-(-1)^{n} J_{2} \Big(S_n^x S_{n+1}^z S_{n+2}^x + S_n^y S_{n+1}^z S_{n+2}^y \Big)
\nonumber\\
&&+(-1)^{n} h_{s} S_n^z \Big]
\label{eq1},
\end{eqnarray}
%
where, $N$ is the size of the system, $h_{s}$ is the magnitude of staggered magnetic field and $\omega$ is the driving frequency.
Here we impose periodic boundary condition and $S_n^\alpha$ are the  spin half operators at the $n$th site, i.e. $S_n^\alpha=\frac{1}{2}\sigma^{\alpha}_{n}; \;\;\; \alpha=\{x,y,z\}$, where $\sigma^{\alpha}$
are Pauli matrices.
The first and second terms in Eq. (\ref{eq1}) describe the time dependent nearest neighbour XY and staggered Dzyaloshinskii-Moriya interactions \cite{Jafari2008}, and the third term is a staggered cluster (three-spin) interaction \cite{Titvinidze}.

The Hamiltonian, Eq. (\ref{eq1}), can be exactly diagonalized by Jordan-Wigner transformation, \cite{} which
transforms spins into spinless fermions, where $c^{\dagger}_{n}$ ($c_{n}$) is the fermion creation (annihilation) operator (see Appendix \ref{A1}).
The crucial step is to define two independent fermions at site $n$, $c_{n-1/2}^{A}=c_{2n-1}$, and $c_{n}^{B}=c_{2n}$.
This can be regarded as splitting the chain having a diatomic unit cell. Introducing the Nambu spinor $\Gamma^{\dagger}_k=(c_{k}^{\dagger B},~c_{k}^{\dagger A})$, the Fourier transformed Hamiltonian can be expressed as sum of independent terms acting in the two-dimensional Hilbert space generated by $k$
%
\begin{eqnarray}
{\cal H}(t)=\sum_{k}\Gamma_{k}^{\dagger}H_{k}(t) \Gamma_{k}
\label{eq2},
\end{eqnarray}
%
The Bloch single particle Hamiltonian $H_{k}(t)$ in Eq. (\ref{eq2}), is $H_{k}(t)=J_{1}\Big[h_{xy}\Big(\cos(\omega t)\sigma^{x}+\sin(\omega t)\sigma^{y}\Big)+h_{z}\sigma^{z}\Big]/2$, where  $h_{xy}(k)=2\cos(k/2)$ and $h_{z}(k)=J_{2}\cos(k)+2h_{s}$.
Eq. (\ref{eq2}) implies that the Hamiltonian of $N$ interacting spins (Eq. (\ref{eq1})) can be mapped to the
sum of $N/2$ noninteracting quasi-spins. In the next section, we deal with the Hamiltonian of noninteracting quasi-spins to show
that quasi-spins transform from nonadiabatic to adiabatic regime by tuning the driving frequency $\omega$.

\subsection{Exact solution of the time-dependent Schr\"{o}dinger equation\label{schrodingerEQ}}

The noninteracting (quasi-spin) Hamiltonian $H_{k}(t)$ is exactly the same as Schwinger-Rabi model \cite{Schwinger1937}
of a spin in the time dependent effective magnetic field $H_{k}(t)=\vec{h}_{k}(t)\cdot\vec{S}$ with
$\vec{h}_{k}(t)=J_{1}(h_{xy}(k)\cos(\omega t),h_{xy}(k)\sin(\omega t),h_{z}(k))$ and $|\vec{h}_{k}|=J_{1}(h^2_{xy}(k)+h^2_{z}(k))^{1/2}$.
In such a case, the polar and azimuthal angles of effective magnetic field, are $\theta_{k}=\arctan(h_{xy}(k)/h_{z}(k))$
and $\varPhi(t)=\omega t$, respectively.

Using the time-dependent Schr\"{o}dinger equation ${\it i}\frac{d}{dt}|\psi_{k}^{\pm}(t)\rangle=H_{k}(t)|\psi_{k}^{\pm}(t)\rangle$ in
the rotating frame given by the periodic unitary transformation $U_{R}(t)=\exp[{\it i}\omega(\mathbb{1}-\sigma^{z})t/2]$,
the time dependent Hamiltonian is transformed to its time-independent form $\mathbb{H}_{k}|\chi^{\pm}_{k}\rangle=E^{\pm}_{k}|\chi^{\pm}_{k}\rangle$ where
%
\begin{equation}
\mathbb{H}_{k}=[h_{xy}(k)\sigma^{x}+(h_{z}(k)-\omega)\sigma^{z}+\omega\mathbb{1}]/2
\label{eq3},
\end{equation}
%
and $|\chi^{\pm}_{k}\rangle=U_{R}^{\dagger}(t)|\psi_{k}^{\pm}(t)\rangle$ (see Appendix \ref{A2}).
For simplicity and without loss of generality we take $J_{1}=1$, and $J_{2}, h_{s}, \omega>0$, henceforth.
The eigenvalues and eigenvectors of the time-independent noninteracting Hamiltonian $\mathbb{H}_{k}$ are
%
\begin{equation}
E^{\pm}_{k}=\frac{\omega}{2}\pm\frac{1}{2} \sqrt{h_{xy}^{2}(k)+(h_{z}(k)-\omega)^{2}}
\label{eq4},
\end{equation}
%
%
\begin{eqnarray}
&&|\chi^{-}_{k}\rangle =\left(
\begin{array}{c}
\cos(\gamma_{k}/2) \\
\sin(\gamma_{k}/2) \\
\end{array}
\right),
\nonumber\\
&&|\chi^{+}_{k}\rangle =\left(
\begin{array}{c}
\sin(\gamma_{k}/2) \\
-\cos(\gamma_{k}/2) \\
\end{array}
\right),
\label{eq5}
\end{eqnarray}
%
where
%
\begin{eqnarray}
\label{eq6}
\gamma_{k}=\arctan\Big[\frac{\sin(\theta_{k})}{\cos(\theta_{k})-\omega/|\vec{h}_{k}|}\Big].
\end{eqnarray}
%
It is worthy to mention that, the time independent extended XY model in the presence of the renormalized staggered magnetic filed
%
\begin{eqnarray}
{\cal H} = \sum_{n=1}^{N}&& J_{1} \Big[\Big(S_n^x S_{n+1}^x + S_n^y S_{n+1}^y \Big)+(-1)^{n} h^{eff}_{s} S_n^z
\nonumber\\
&&+(-1)^{n} J_{2} \Big(S_n^x S_{n+1}^z S_{n+2}^x + S_n^y S_{n+1}^z S_{n+2}^y \Big)\Big]
\label{eq7},
\end{eqnarray}
%
with $h^{eff}_{s}=h_{s}-\omega/2$, results the noninteracting (quasi-spin) Hamiltonian $\mathbb{H}_{k}$, the same as Eq. (\ref{eq3}), (apart from an additive constant), which leads to eigenvectors and eigenvalues given by Eqs. (\ref{eq4})-(\ref{eq5}).
It is clear that the ground state is separated from the excited state by the energy gap
$\Delta_{k}=|E^{+}_{k}-E^{-}_{k}|$, which vanishes at Brillouin zone boundary $k_{c}=\pm\pi$ and $h^{eff}_{s}=h_{s}-\omega/2=J_{2}/2$. So, a QPT occurs at
$h^{eff}_{s}=J_{2}/2$, where the system transforms from the Spin Liquid phase ($h^{eff}_{s}<J_{2}/2$)
to the long-range ordered antiferromagnetic phase ($h^{eff}_{s}>J_{2}/2$).
%
\begin{figure}
\centerline{\includegraphics[width=1\columnwidth]{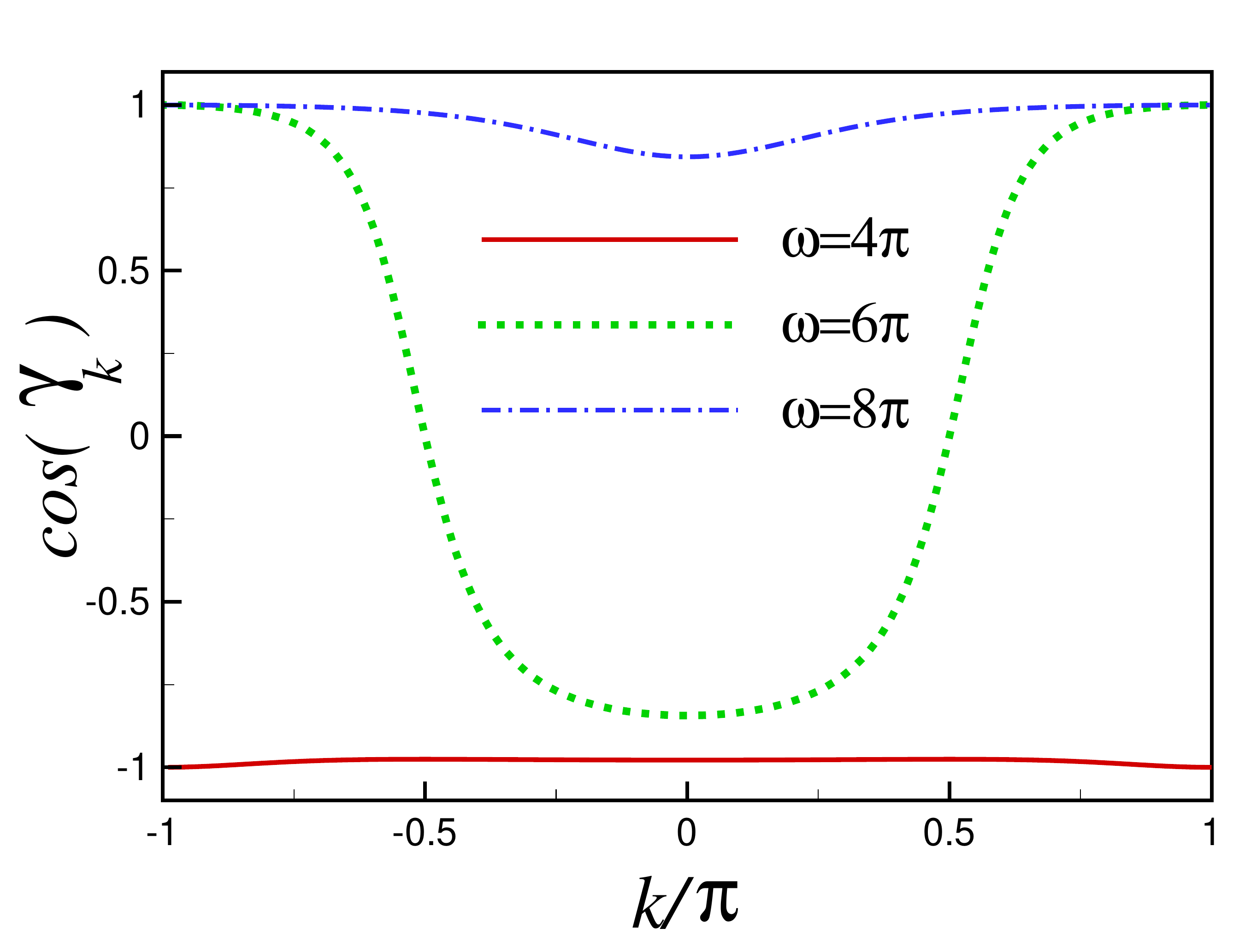}}
\caption{(Color online) Variation of $\cos(\gamma_k)$ versus $k/\pi$ for $J_2=\pi, h_{s}=3\pi$ and
$\omega=4\pi$ (solid line), $\omega=6\pi$ (dotted line) and $\omega=8\pi$
(dash-dotted line).}
\label{fig1}
\end{figure}
%

\subsection{Topological transition from adiabatic to nonadiabatic regime\label{diabatictoadiabatic}}
An adiabatic cyclic process associates with slowly changing the periodic parameters driving a physical system. In such a case,
the system returns to its initial state after a cycle. However, a quantum state may gain a phase factor that can be given as the sum of a dynamical phase that depends on the Hamiltonian parameters and an extra term of geometrical origin. The latter is known as the Berry phase and can be determined from the geometric properties of the path expressed by the driving parameters in the parameter space of the Hamiltonian \cite{Berry1984}. Nonadiabatic cyclic processes may also produce geometric phases that are smaller than adiabatic Berry phases but with similar geometric characteristics \cite{Aharonov1987,Bohmbook,JAFARI20133279}.

Consider the instantaneous non-degenerate eigenstates $|m,\textbf{R}\rangle$ of the system for a given set of parameters \textbf{R} in $H(\textbf{R})$, i.e. $H(\textbf{R})|m,\textbf{R}\rangle=E_{m}(\textbf{R})|m,\textbf{R}\rangle$, where $E_{m}(\textbf{R})$ is the corresponding eigenvalues.
Hence, the Berry phase is expressed as $\Upsilon_{m}=\oint A^{m}(\textbf{R})$, where $A^{m}(\textbf{R})$ is the so-called Berry connection expressed as a (local) differential one-form \cite{Bohmbook}
%
\begin{equation}
A^{m}(\textbf{R})=A^{m}_{\upsilon}dR^{\upsilon}=i\langle m,\textbf{R}|\frac{\partial}{\partial R^{\upsilon}}|m,\textbf{R}\rangle dR^{\upsilon}
\label{eq8}
\end{equation}
%
Using Stokes theorem, we arrive at $\Upsilon_{m}=\int_{S} F^{m}(\textbf{R})$, where $F^{m}(\textbf{R})$ is the Berry curvature
two-form given by \cite{Bohmbook}
%
\begin{equation}
F^{m}(\textbf{R})=\frac{\partial A^{m}_{\mu}}{\partial R^{\upsilon}}dR^{\upsilon}\wedge dR^{\mu}=dA^{m}.
\label{eq9}
\end{equation}
%
Here, $\wedge$ is the antisymmetric wedge product and $dA^{m}$ stands for the exterior derivative of the
Berry connection one-form $A^{m}$. The Chern number, which is given by the integral of the Berry curvature
over the whole parameter space, encodes the information of a topological transition between two different driving
regimes (see Appendix \ref{A4}). In the adiabatic regime, a spin can adapt to the variation of effective magnetic
field and will remain in its instantaneous eigenstate during the slow evolution. In contrast, in the nonadiabatic regime,
the spin cannot align with the magnetic field and hence is exposed to an average
magnetic field \cite{gomez2012topological,zener1932non,betthausen2012spin,leon2014dynamical}.
%
\begin{figure*}
\begin{minipage}{\linewidth}
\centerline{\includegraphics[width=0.31\linewidth]{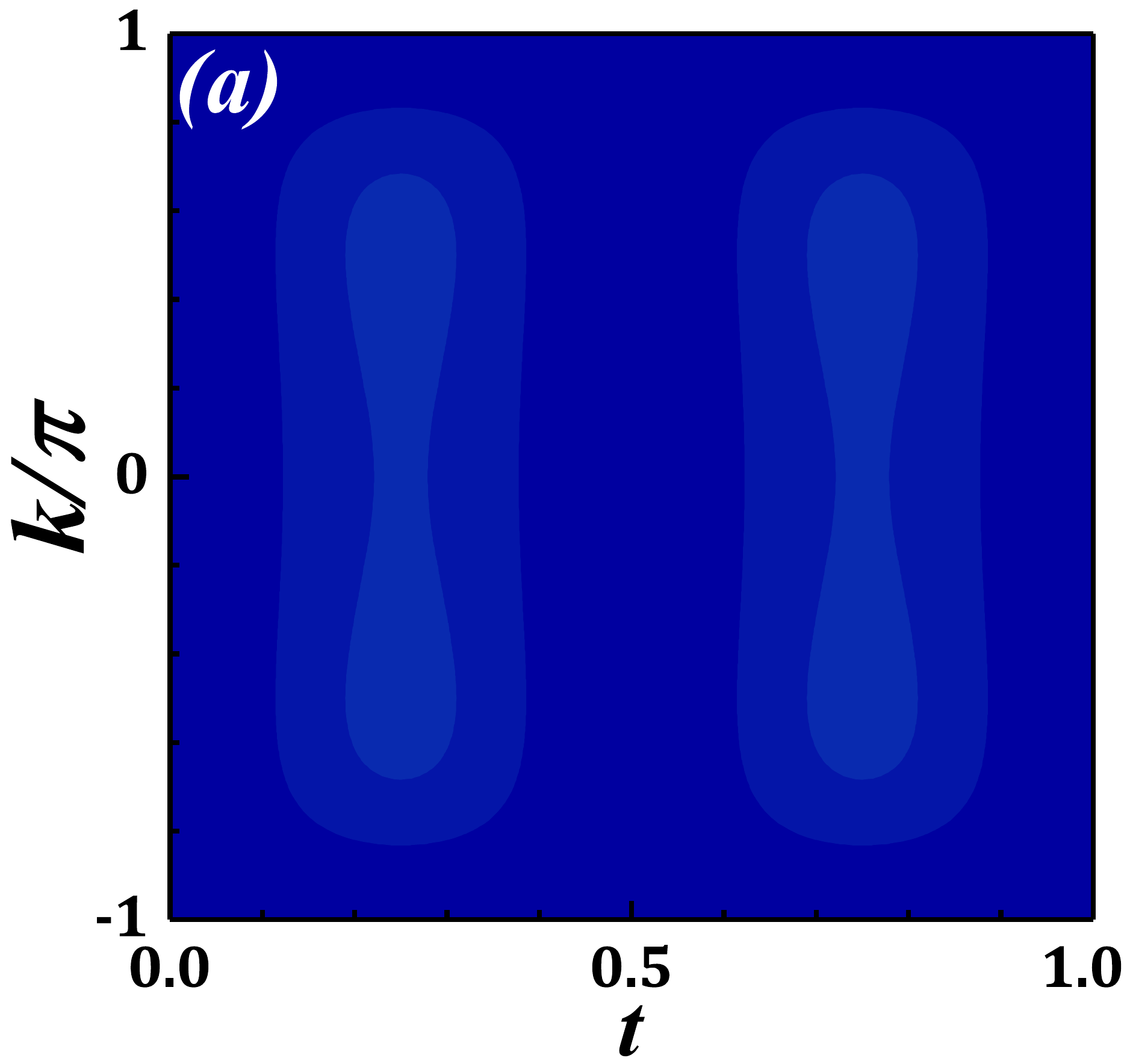}
\includegraphics[width=0.31\linewidth]{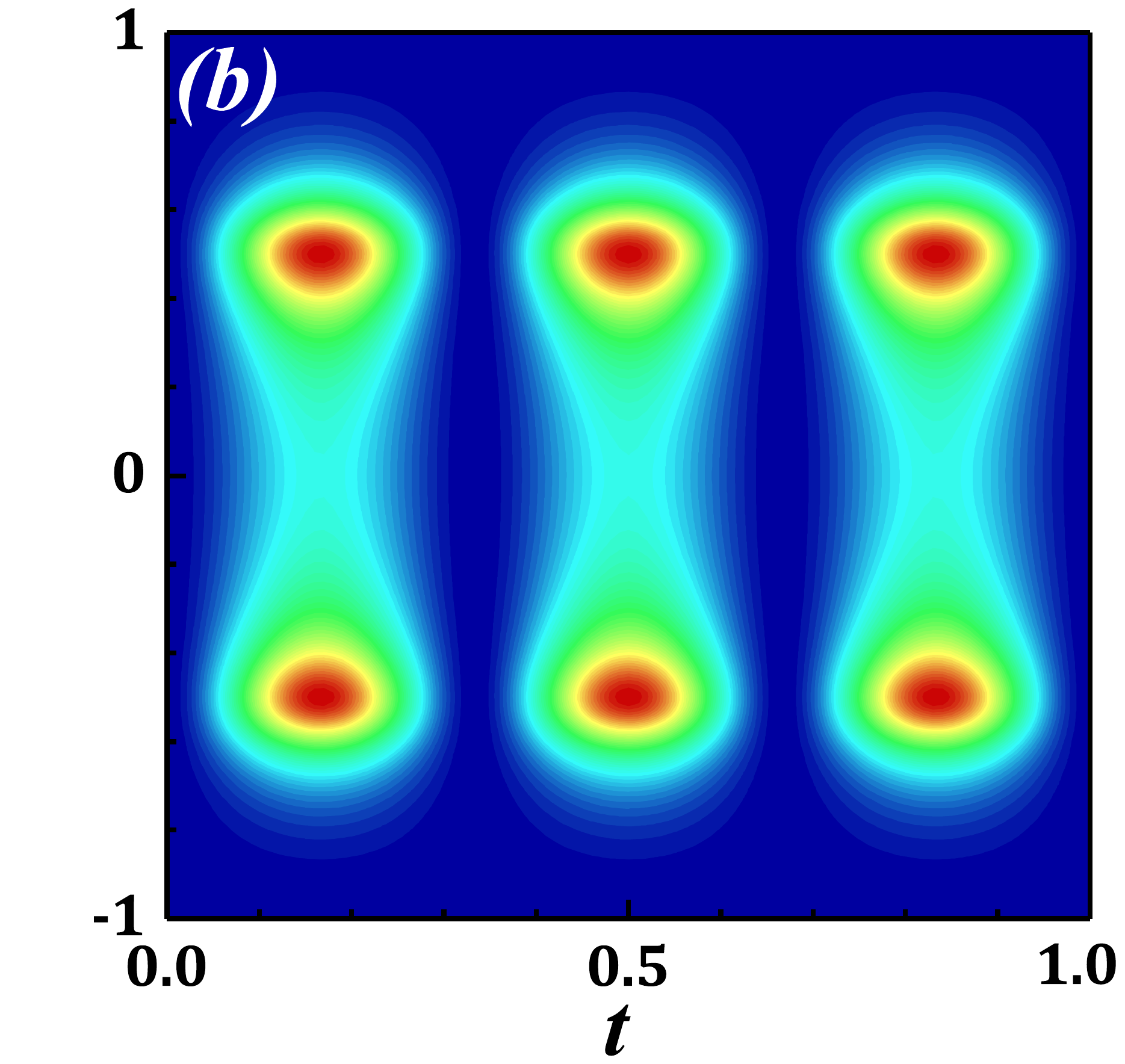}
\includegraphics[width=0.37\linewidth]{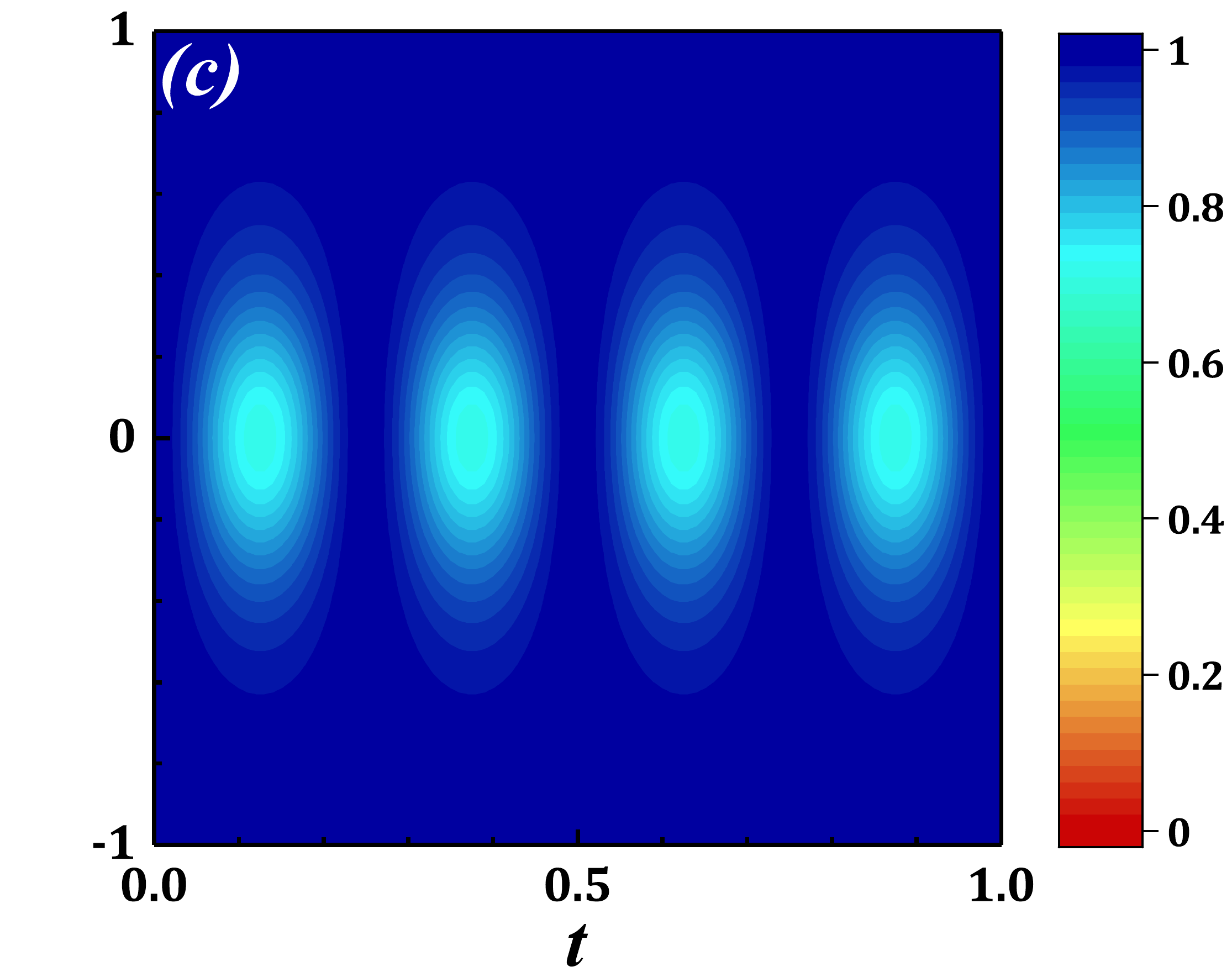}}
\centering
\end{minipage}
\begin{minipage}{\linewidth}
\centerline{\includegraphics[width=0.33\linewidth]{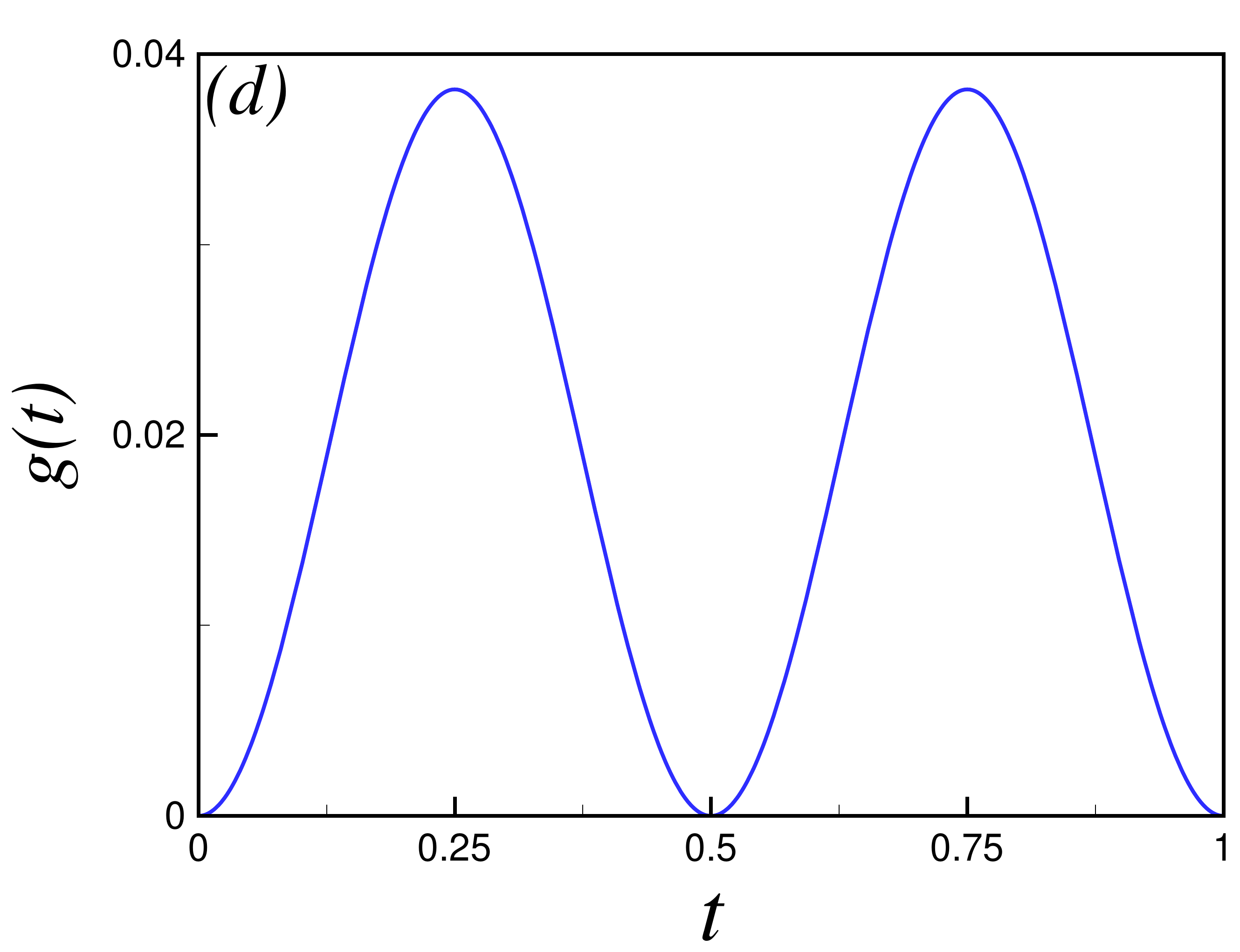}
\includegraphics[width=0.33\linewidth]{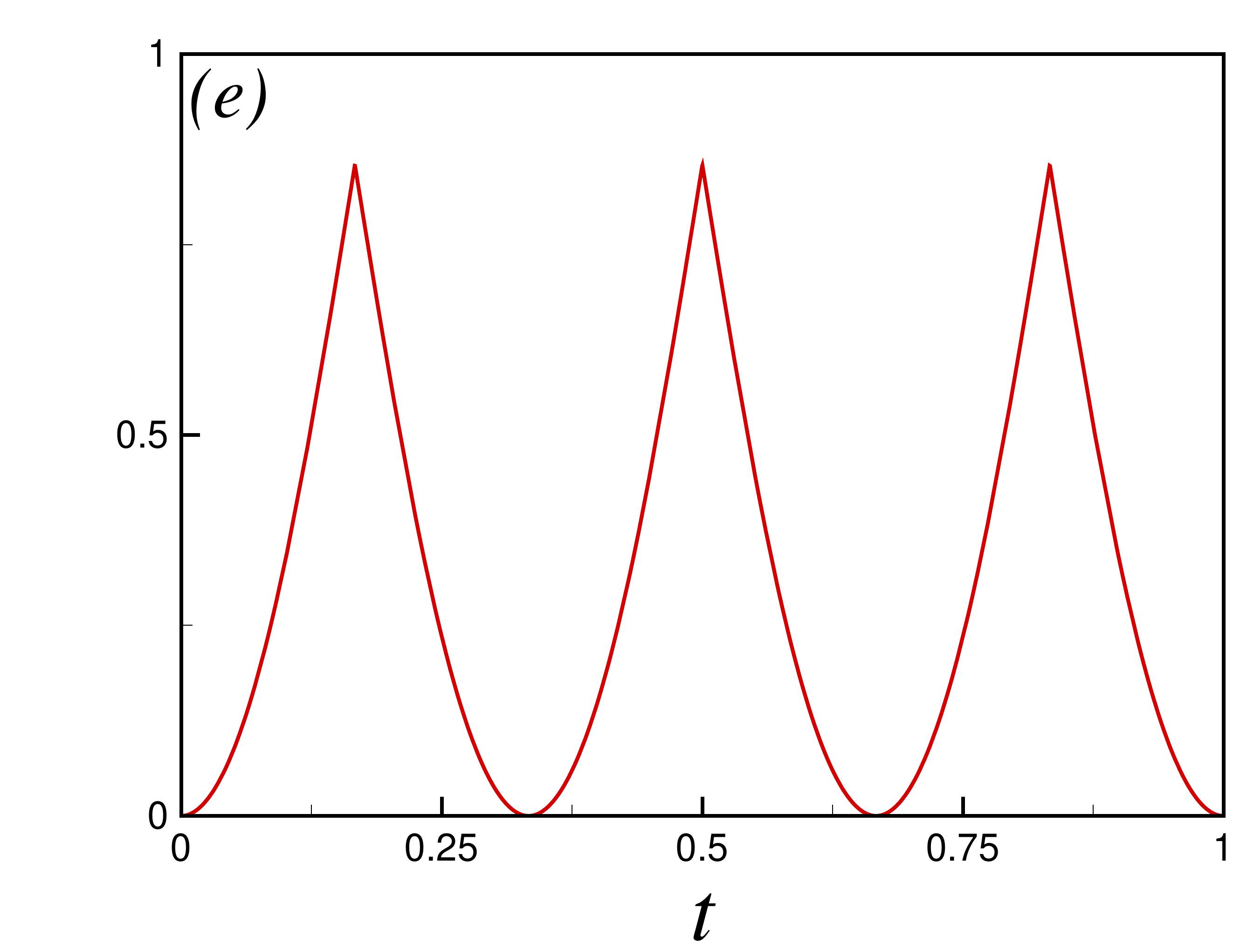}
\includegraphics[width=0.33\linewidth]{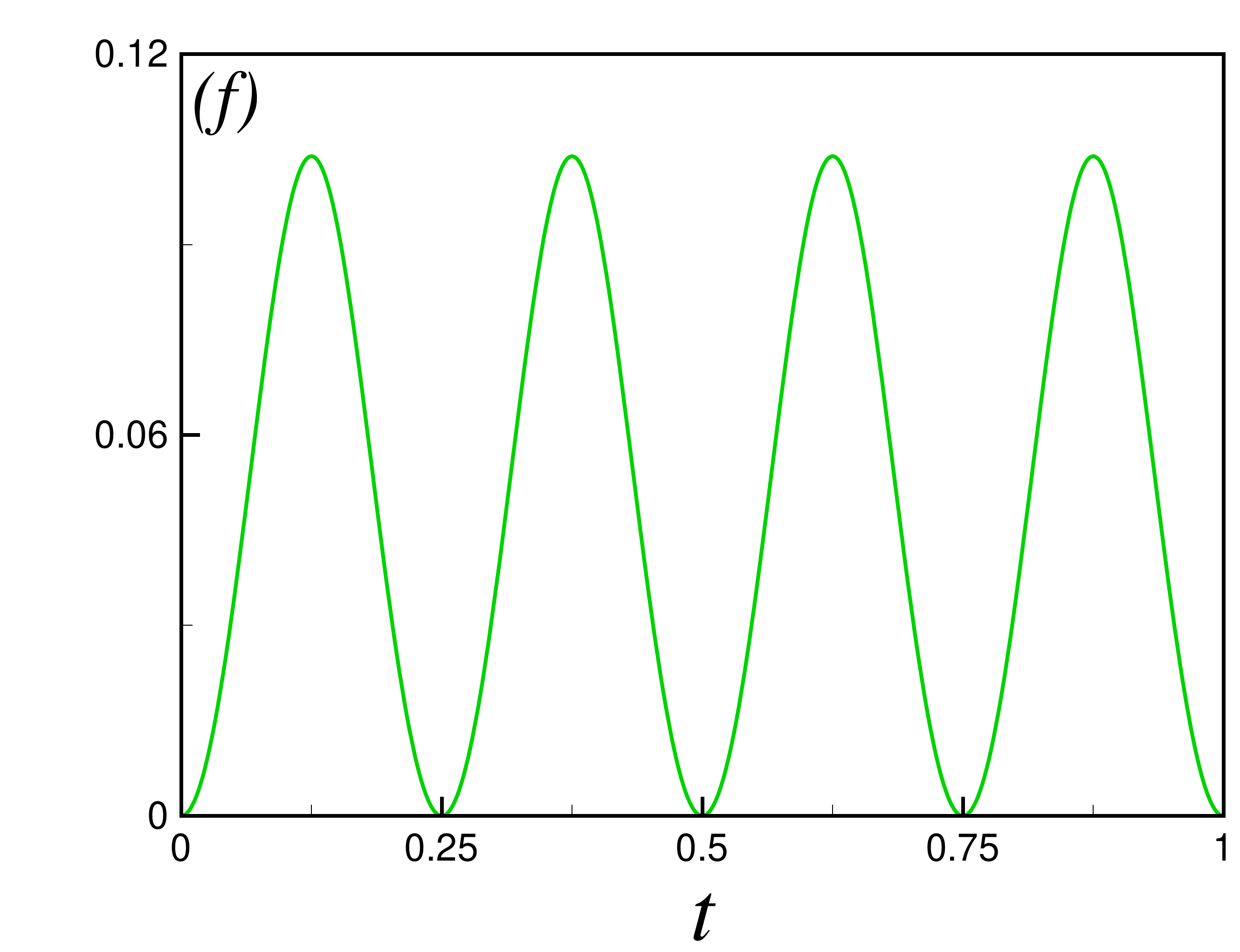}}
\centering
\end{minipage}
\caption{(Color online) The density plot of Loschmidt echo $|{\cal L}_{k}(t)|^{2}$
as a function of time $t$ and $k$, for  (a) $\omega=4\pi$
(b) $\omega=6\pi$ and (c) $\omega=8\pi$. The rate function of Loschmidt amplitude
versus time $t$ for (d) $\omega=4\pi$
(e) $\omega=6\pi$ and (f) $\omega=8\pi$.
In all plots we take $J_{2}=\pi$, $h_{s}=3\pi$.
}
\label{fig2}
\end{figure*}
%
Implementing the exact Floquet states, we obtain
the Berry curvature, which gives an exact expression for the Chern number ($C$) (see Appendix \ref{A4}).
%
\begin{equation}
C=\Theta (1-\frac{\omega}{\sqrt{h_{xy}^2(k)+h_{z}^2(k)}})=\Theta (1-\frac{\omega}{|\vec{h}_{k}|}),
\label{eq10}
\end{equation}
%
where $\Theta(x)$ is the Heaviside step function.

According to the calculated Chern number (Eq. (\ref{eq10})), in the adiabatic regime $\omega<|\vec{h}_{k}|$  we get $C=1$, where the quasi-spin is able to follow the evolution of magnetic field $\vec{h}_{k}$, leading to the Rabi oscillation \cite{Schwinger1937}. However, in the nonadiabatic regime $\omega>|\vec{h}_{k}|$ ($C=0$), which corresponds to the Landau Zener transition \cite{zener1932non}, the spin can not follow the effective magnetic field and is only subjected to the average field \cite{betthausen2012spin,leon2014dynamical}. To simply explain the transition from adiabatic to nonadiabatic regime, let us to revisit the definition of angle $\gamma_{k}$ in Eq. (\ref{eq6}).
To get oscillations on the quasi-spin,
$\gamma_{k}$ needs to be varied from
$0$ to $\pi$, i.e., $\gamma_{k}\in [0,\pi]$. This is possible only if $\omega<|\vec{h}_{k}|$, such that the denominator of Eq. (\ref{eq6}) can become zero leading the argument runs from $(-\infty,\infty)$. In turn, it is required that the driving frequency ranges from $|J_{2}-2h_{s}|$ to $\sqrt{(J_{2}+2h_{s})^2+4}$, i.e., $|J_{2}-2h_{s}|\leq\omega\leq\sqrt{(J_{2}+2h_{s})^2+4}$. In Fig. \ref{fig1}, variation of $\cos(\gamma_{k})$ has been plotted versus $k/\pi$ at $J_{2}=\pi$, $h_s=3\pi$ and for different values of $\omega$. As seen, $\gamma_{k}$ changes from $0$ to nearly $\pi$ (dotted line), where the driving frequency is $\omega=6\pi$ and forces system to
evolve adiabatically. On the other hand, in the nonadiabatic regime ($\omega=4\pi, 8\pi$), $\cos(\gamma_{k})$ is roughly constant, which means that quasi-spins cannot align with the effective magnetic field.

\section{Dynamical Phase Transition\label{DQPTs}}
As mentioned in the Introduction, there has been recently a renewed focus in the study of DQPTs, probing nonanalyticities
of the dynamical free energy of a quenched system in both pure and mixed states \cite{heyl2018dynamical,heyl2017dynamical,bhattacharya2017mixed}.
The DQPTs notion emanates from the resemblance between the canonical partition function of an equilibrium system
$Z(\beta)=Tr e^{-\beta{\cal H}}$ and the quantum boundary partition function $Z(z)=\langle\Psi|e^{-z{\cal H}}|\Psi\rangle$ \cite{LeClair,Piroli}. When $z=it$ the quantum boundary partition function corresponds to the Loschmidt amplitude
${\cal L}(t)=\langle\Psi(\lambda_{1})|e^{-i{\cal H}(\lambda_{2})t}|\Psi(\lambda_{1})\rangle$.
In such a case, LA is the overlap amplitude of the initial quantum state $|\Psi(\lambda_{1})\rangle$ with its time evolved state
under the post-quenched Hamiltonian ${\cal H}(\lambda_{2})$.
The DPTs are defined by sharp nonanalyticities in the rate function of the Loschmidt echo (LE)-square
given by \cite{Pollmann,heyl2013dynamical,andraschko2014dynamical,sharma2015quenches},
%
\begin{eqnarray}
\no
g(t)=-\frac{1}{2\pi}\int_{-\pi}^{\pi} dk \ln|{\cal L}_{k}(t)|^{2}.
\end{eqnarray}
%
In addition, these nonanalyticities are signaled by the zeros of $Z(z)$, known as Fisher zeros \cite{heyl2013dynamical,heyl2018dynamical}.
Here, we investigate the Floquet DQPTs in the proposed periodically time dependent Hamiltonian Eq. (\ref{eq1}) to explore characteristics of DQPTs in the quantum Floquet systems.

\subsection{Pure state dynamical phase transition\label{pureDQPT}}
According to the results of section \ref{schrodingerEQ} the time-evolved ground state
$|\psi^{-}_{k}(t)\rangle$ of the quasi-spin Hamiltonian $H_{k}(t)$, is given by
%
{\small
\begin{eqnarray}
|\psi^{-}_{k}(t)\rangle&=&U_{R}(t)e^{-iE^{-}_{k}t}|\chi^{-}_{k}\rangle
=e^{-iE^{-}_{k}t}e^{i\omega(\mathbb{1}-\sigma^{z})t/2}|\chi^{-}_{k}\rangle.
\label{eq11}
\end{eqnarray}
}
%
Due to the decoupling of different momentum sectors, the initial and time-evolved ground states of
the original Hamiltonian exhibit a factorization property that is expressed by
%
\begin{eqnarray}
&&|\psi^{-}(t)\rangle=\Pi_{k}|\psi^{-}_{k}(t)\rangle=\Pi_{k} e^{-iE^{-}_{k}t}U_{R}(t)|\chi^{-}_{k}\rangle,
\nonumber\\
&&|\psi^{-}(t=0)\rangle=\Pi_{k}|\chi^{-}_{k}\rangle.
\label{eq12}
\end{eqnarray}
%

It is straightforward to show that the LA corresponding to the ground state of the proposed
model is given by
%
\bea
\no
{\cal L}(t)&=&\langle\psi^{-}(0)|\psi^{-}(t)\rangle=\Pi_{k}{\cal L}_{k}(t),\\
\no
{\cal L}_{k}(t)&=&\langle\psi^{-}_{k}(0)|\psi^{-}_{k}(t)\rangle
=e^{-iE^{-}_{k}t}\langle\chi^{-}_{k}|U_{R}(t)|\chi^{-}_{k}\rangle\\
\label{eq13}
&=&e^{-iE^{-}_{k}t}\Big[\cos^2(\frac{\gamma_{k}}{2})+\sin^2(\frac{\gamma_{k}}{2})e^{i\omega t}\Big].
\eea
%
%
\begin{figure*}
\centerline{\includegraphics[width=0.325\linewidth]{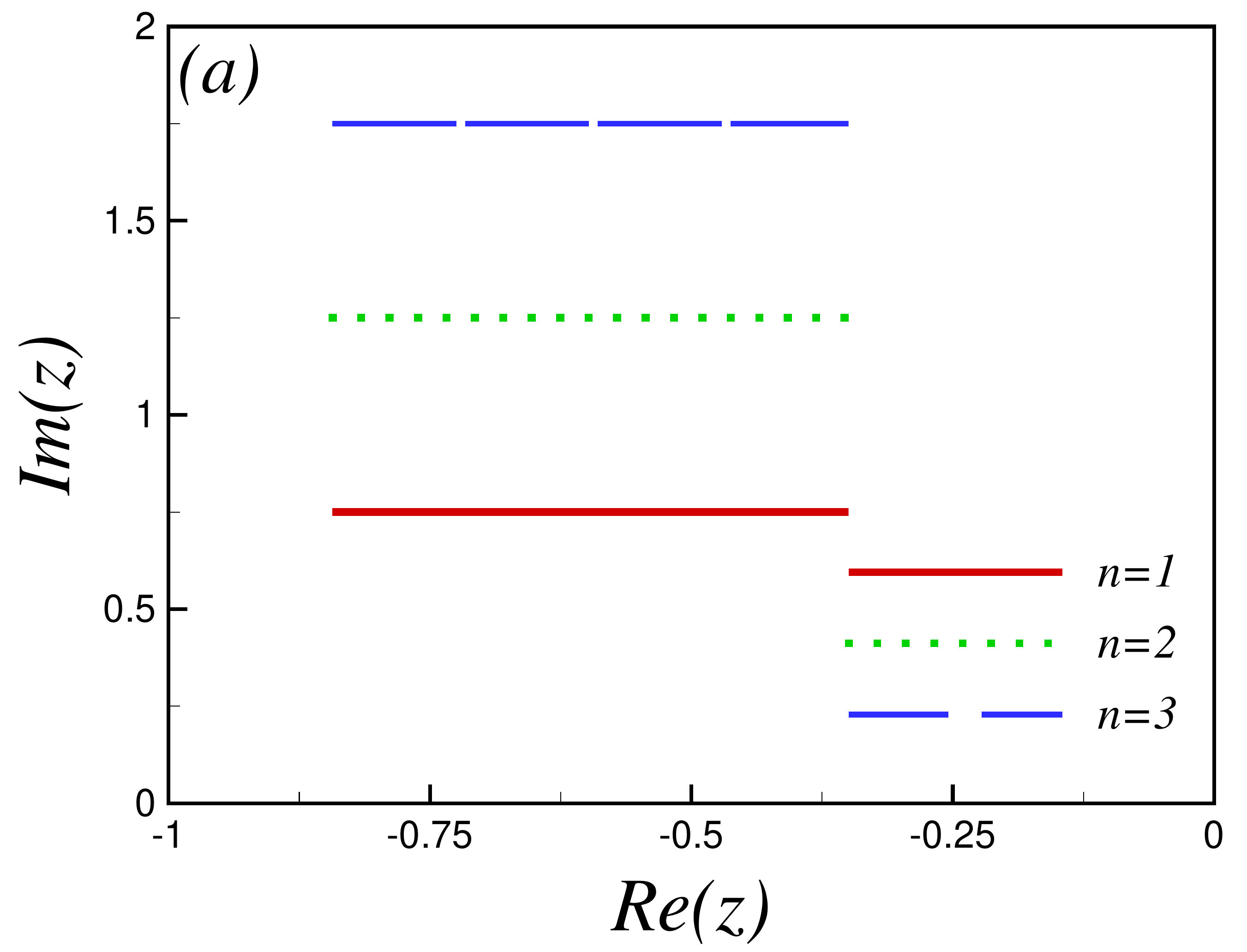}
\includegraphics[width=0.325\linewidth]{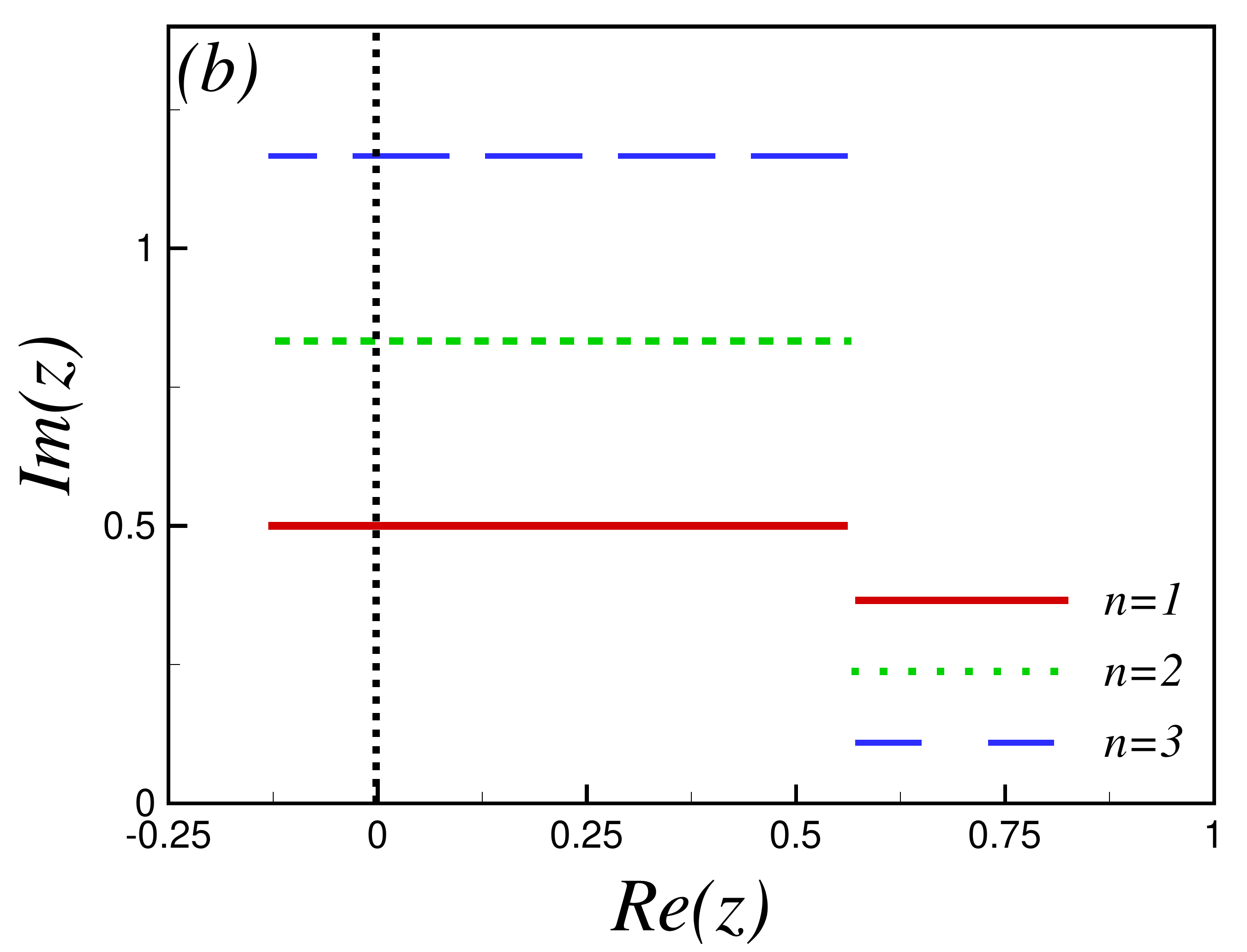}
\includegraphics[width=0.325\linewidth]{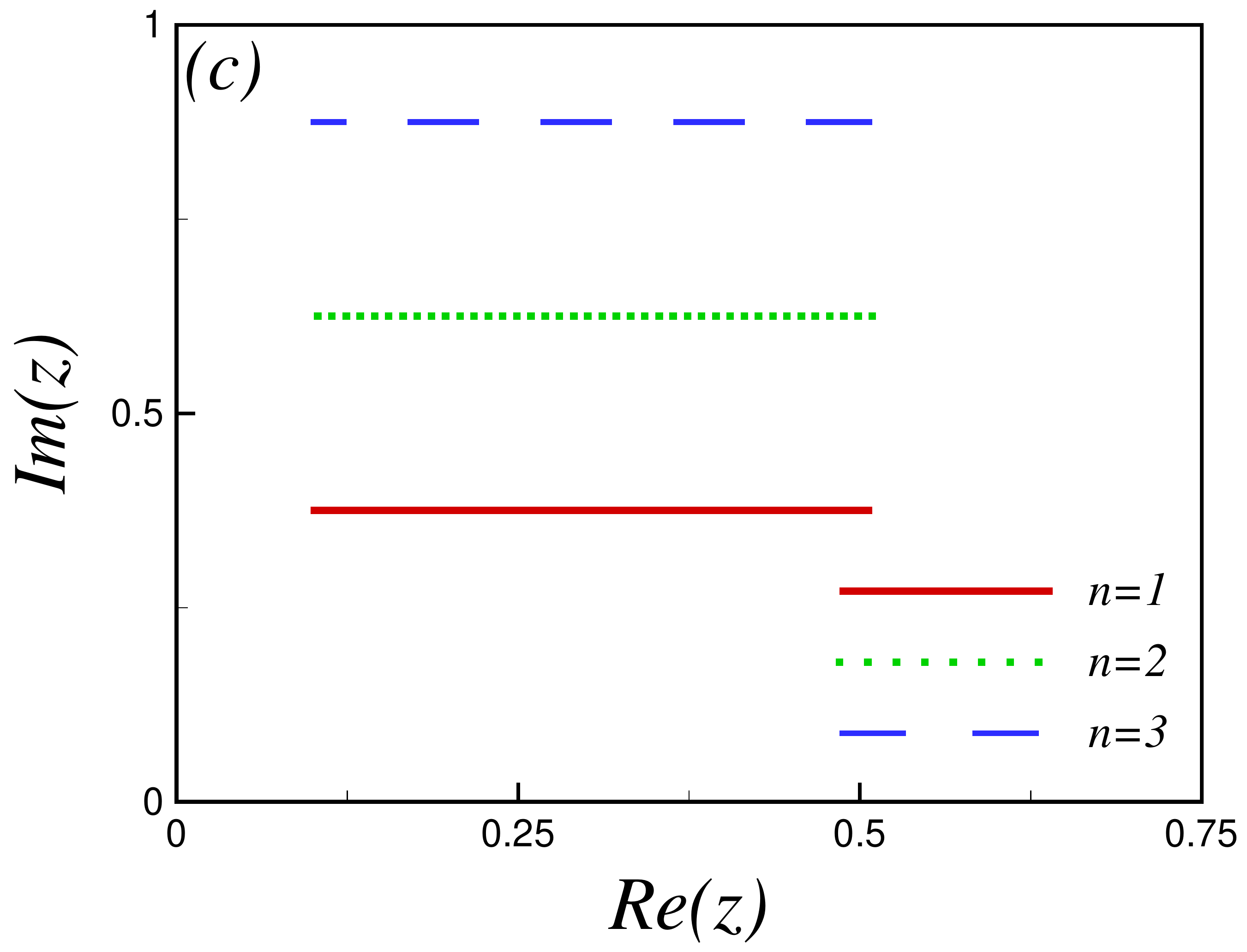}}
\caption{(Color online) Lines of Fisher zeros for $J_{2}=\pi$, $h_{s}=3\pi$, and (a) $\omega=4\pi$,
(b) $\omega=6\pi$ and (c) $\omega=8\pi$.}
\label{fig3}
\end{figure*}
%
%
\begin{figure*}
\centerline{\includegraphics[width=0.33\linewidth]{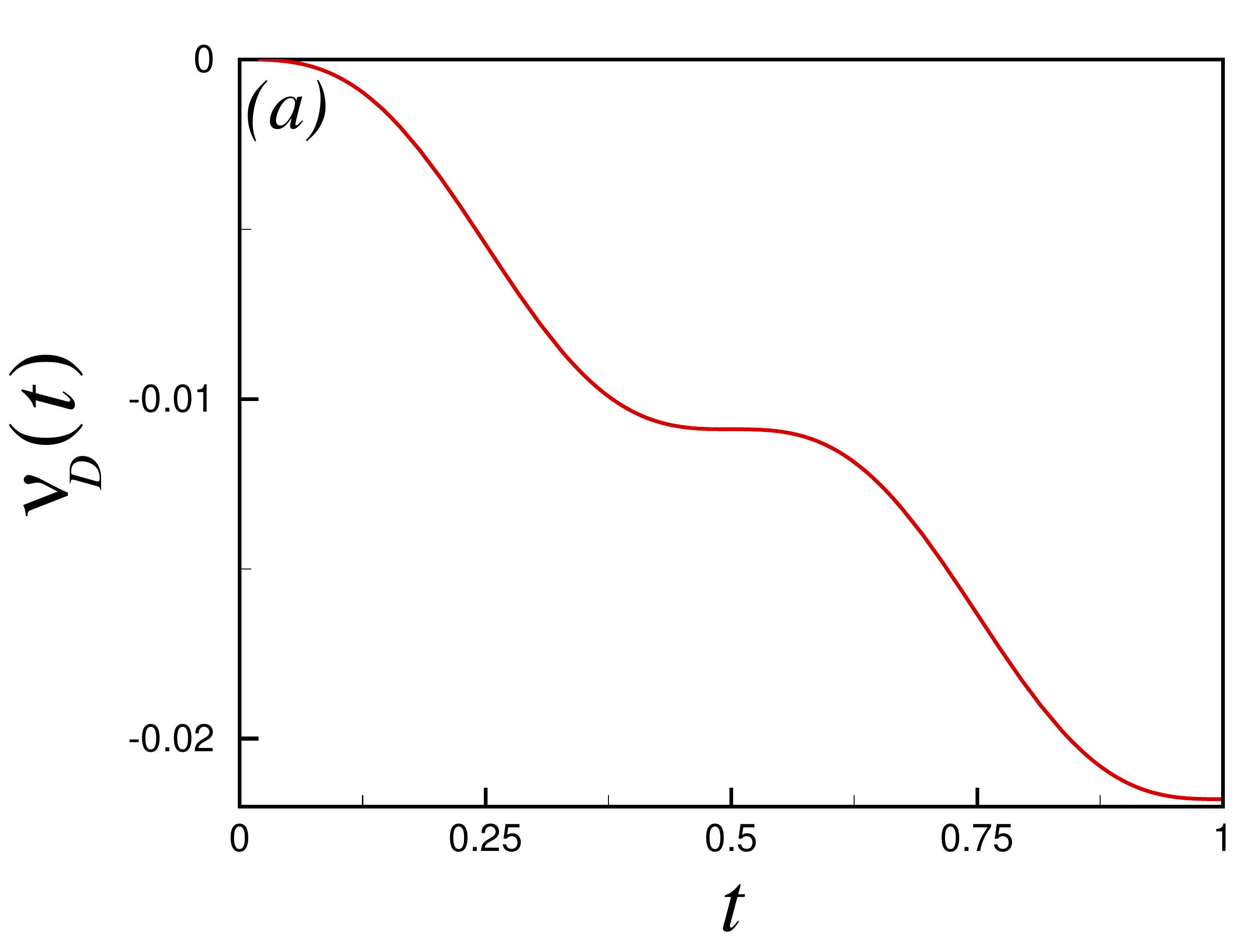}
\includegraphics[width=0.33\linewidth]{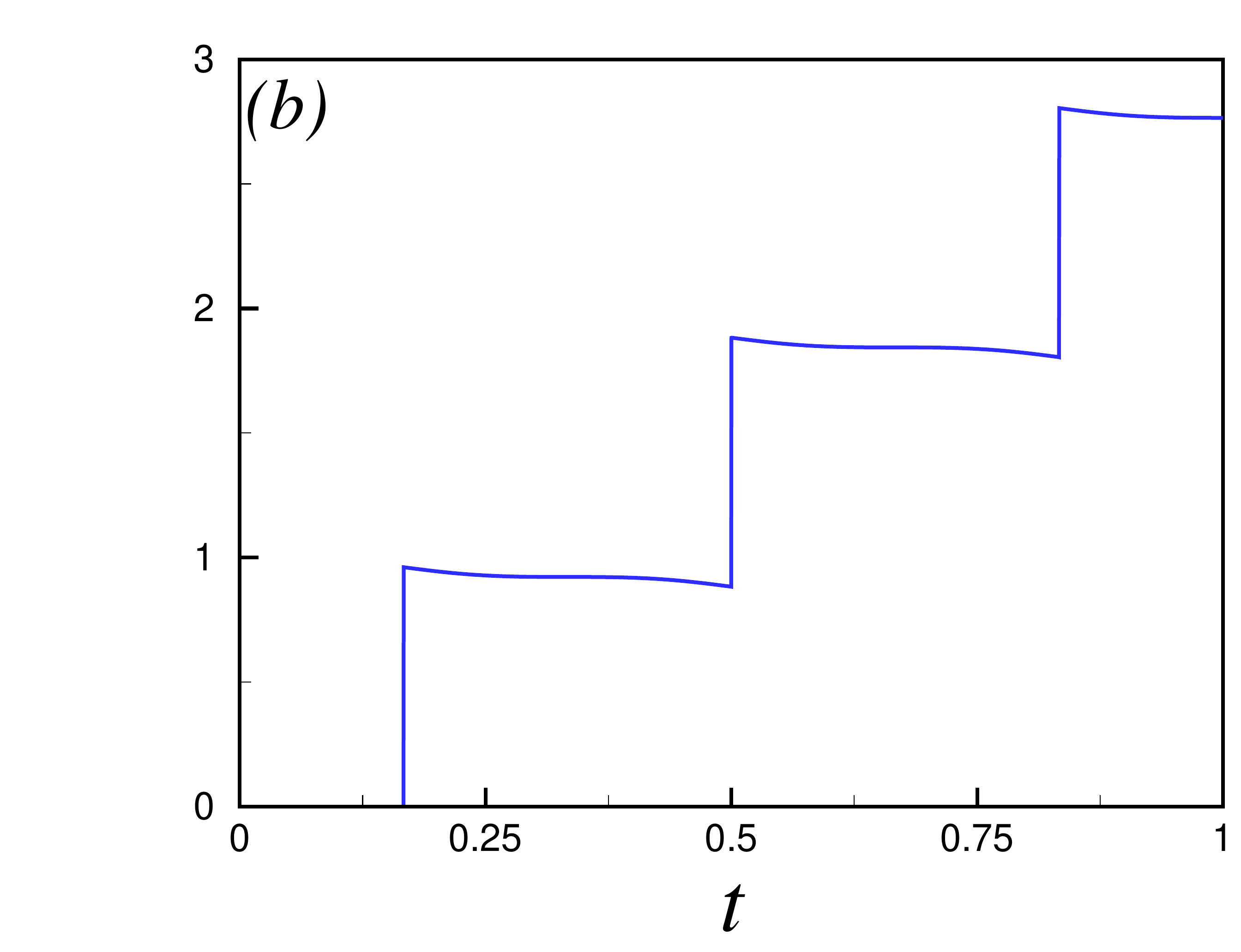}
\includegraphics[width=0.33\linewidth]{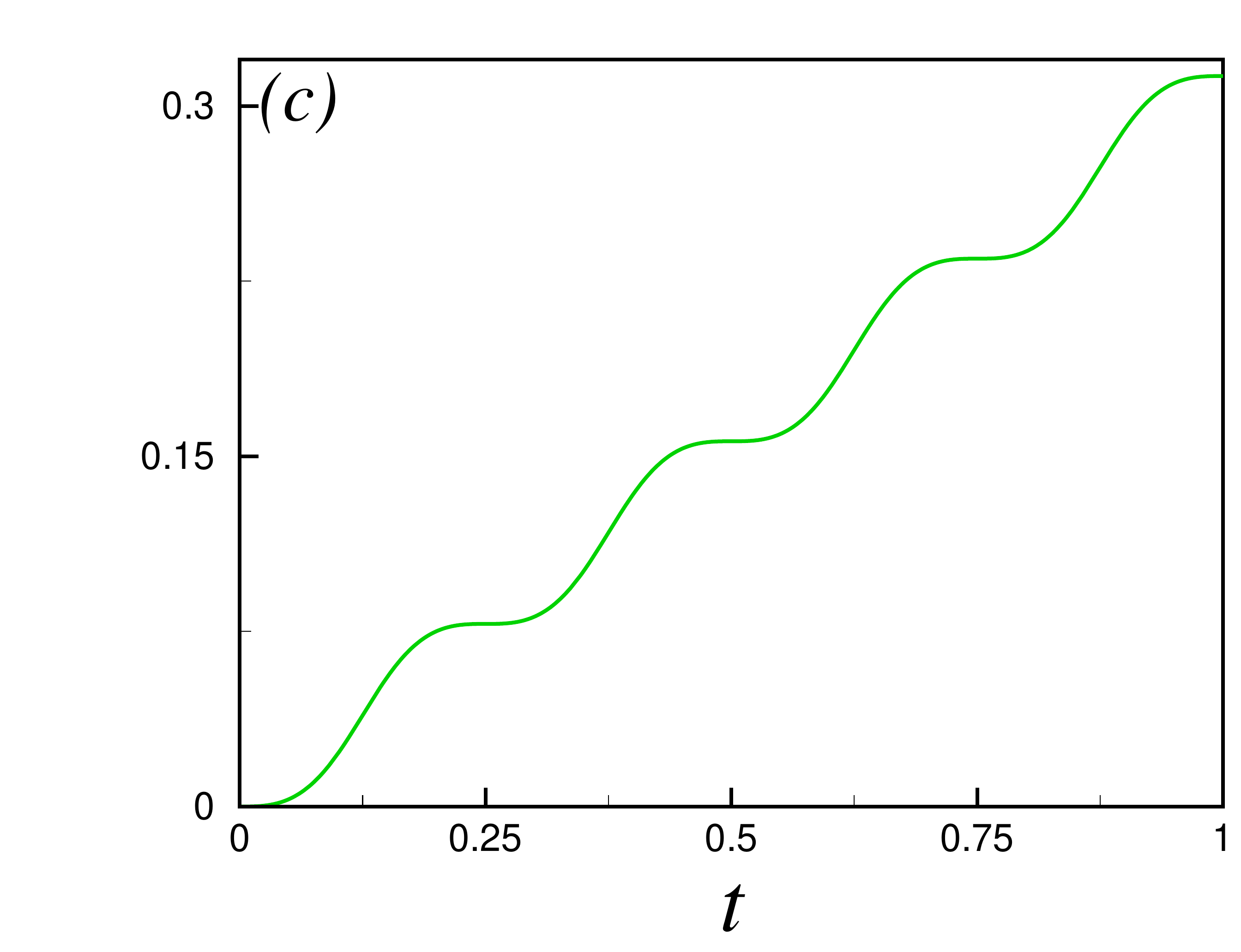}}
\caption{(Color online) The dynamical topological order parameter as a function of time for
$J_{2}=\pi$, $h_{s}=3\pi$ and (a) $\omega=4\pi$, (b) $\omega=6\pi$ and (c) $\omega=8\pi$.}
\label{fig4}
\end{figure*}
%
The density plot of LE and the rate function of LE have been shown in Figs. (\ref{fig2})(a)-(f).
It can be clearly seen that, in the adiabatic regime (Fig. \ref{fig2}(b)) there exist critical points $k^{\ast}$ and $t^{\ast}$, where ${\cal L}_{k^{\ast}}(t^{\ast})$ becomes zero. In contrast, there is no such critical point in nonadiabatic regime (Figs. \ref{fig2}(a) and \ref{fig2}(c)).
Consequently, the nonanalyticities in the rate function of the LA and DQPT occur for the driving frequency, at which the system evolves adiabatically (Fig. \ref{fig2}(e)).

The real time instances at which the DQPT appears is exactly equal to the time instances at which at least one factor
in LA becomes zero i.e., ${\cal L}_{k^{\ast}}(t^{\ast})=0$.
According to Eq. (\ref{eq13}), we find that DQPT happens only whenever there is a mode $k^{\ast}$, which satisfies $J_{2}\cos(k^{\ast})+2h_{s}-\omega=0$,
that leads to $|J_{2}-2h_{s}|<\omega<J_{2}+2h_{s}$.
Since $J_{2}+2h_{s}<\sqrt{4+(J_{2}+2h_{s})^{2}}$, we come to conclude that the nonanalyticities in the rate function of LA can only exist whenever the system evolves adiabatically.
In other words, the minimum required driving frequency for the emergence of Floquet DQPT is equal to the threshold frequency needed for transition from nonadiabatic to adiabatic regime.
Consequently, LA shows a periodic sequence of real-time nonanalyticities in adiabatic regime at
%
\begin{equation}
t^{\ast}_{n}=(2n+1)\frac{\pi}{\omega}=(n+\frac{1}{2}) t^{\ast},~~n{\cal{2}}\mathbb{Z}^+
\label{eq14},
\end{equation}
%
with the period $T_{p}=t^{\ast}=2\pi/\omega$. This result is in agreement with the numerical simulation shown in Figs. \ref{fig2}(b) and \ref{fig2}(e). Furthermore, in contrast to DQPT in the conventional quench mechanism, in which the height of rate function singularity decays with time, the height of cusps in Floquet DQPT does not decay with time.

As mentioned, nonanalyticities of the LA rate function are
accompanied by Fisher zeros of the boundary partition function. The boundary partition function of our model is given by
%
\begin{equation}
{\cal L}_{k}(z)=e^{-E^{-}_{k}z}\Big[\cos^2(\frac{\gamma_{k}}{2})+\sin^2(\frac{\gamma_{k}}{2})e^{\omega z}\Big]
\label{eq15}.
\end{equation}
%
Zeros of Eq. (\ref{eq15}) coalesce in the thermodynamic limit to the family of lines labeled by a number $n \in \mathbb{Z}$:
%
\begin{eqnarray}
z_{n}=\frac{1}{\omega}\Big[\ln(\frac{1+\cos(\gamma_{k})}{1-\cos(\gamma_{k})})
+i(2n+1)\pi\Big].
\label{eq16}
\end{eqnarray}
%
%
\begin{figure*}
\begin{minipage}{\linewidth}
\centerline{\includegraphics[width=0.31\linewidth]{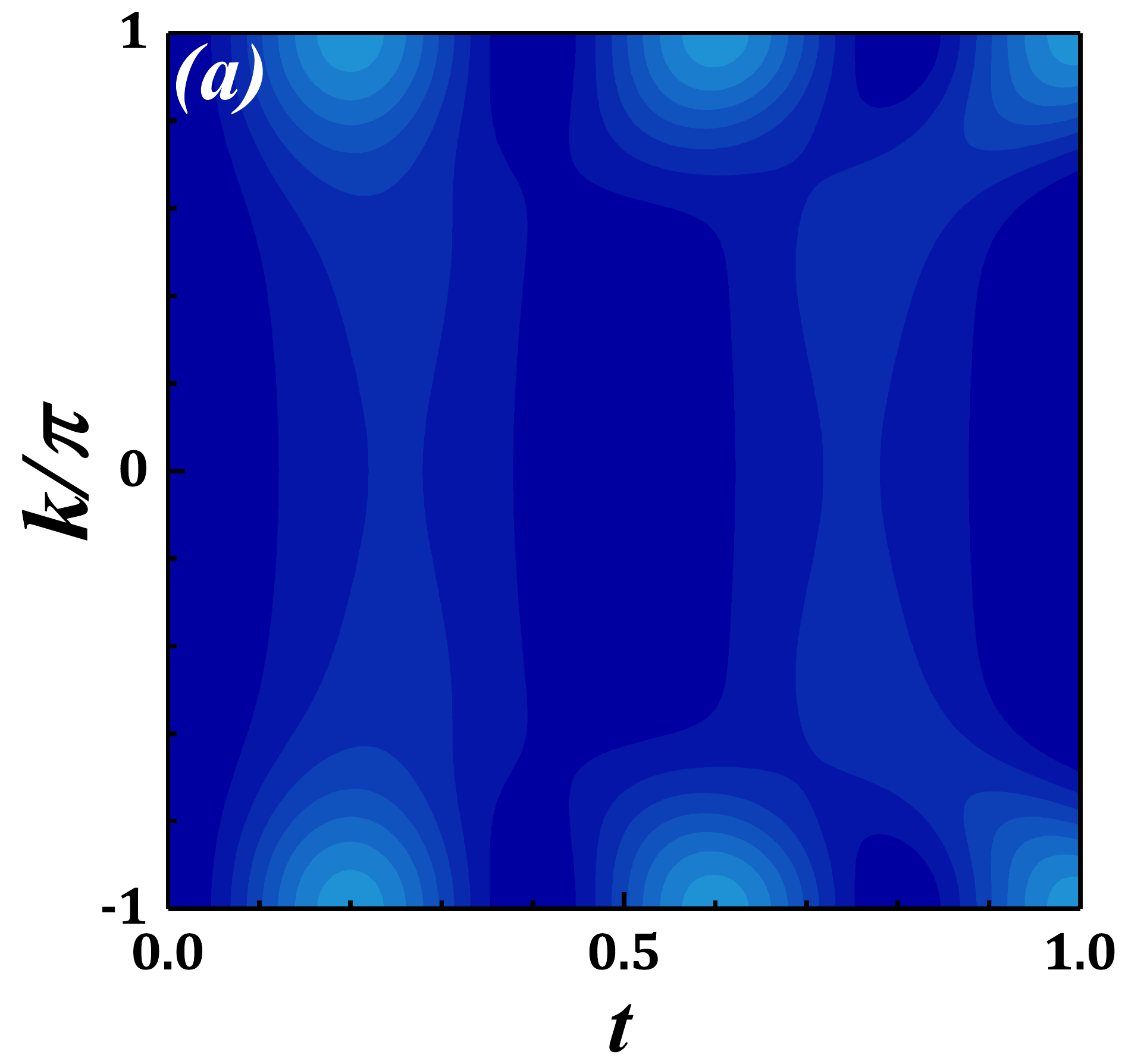}
\includegraphics[width=0.31\linewidth]{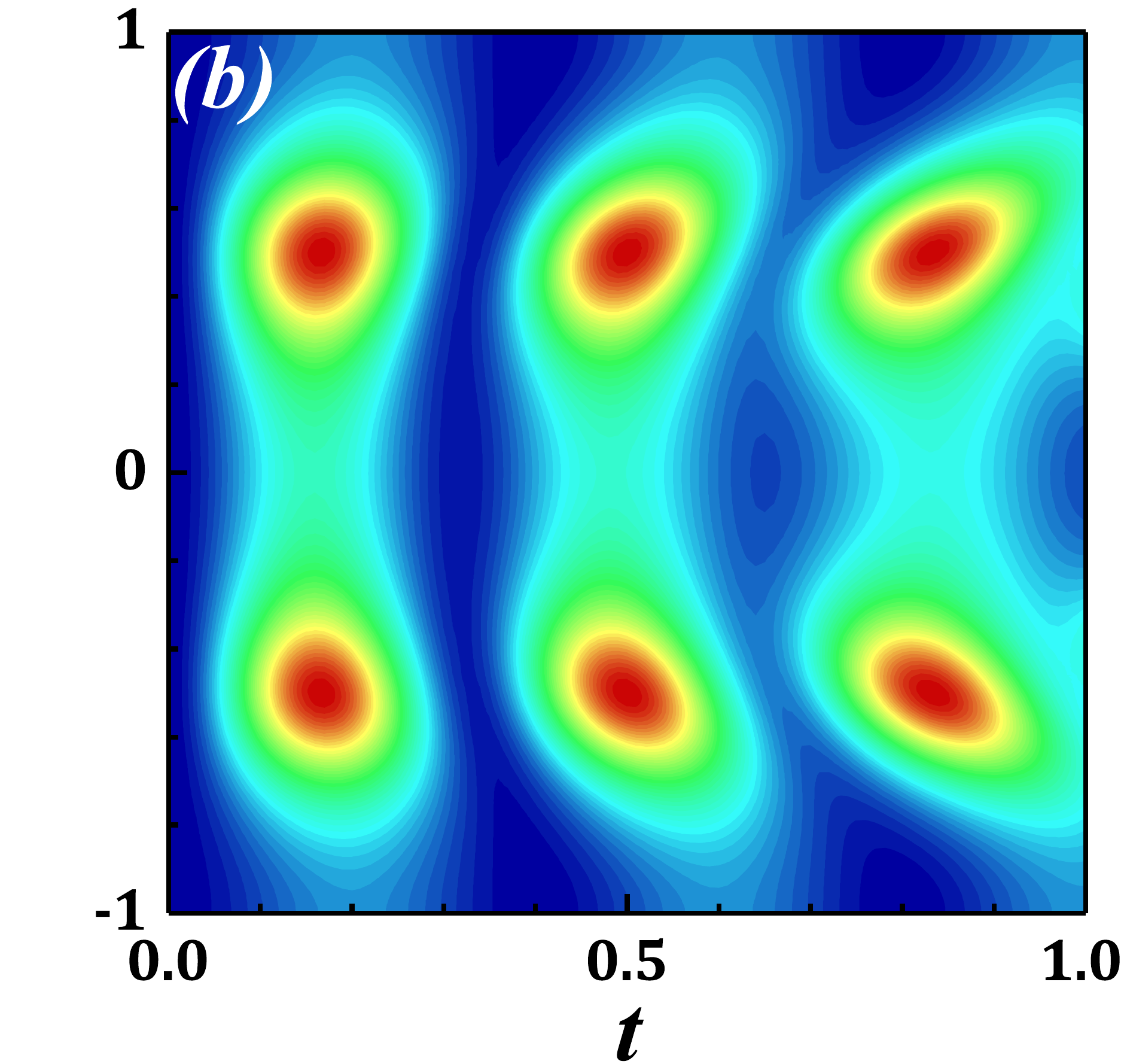}
\includegraphics[width=0.37\linewidth]{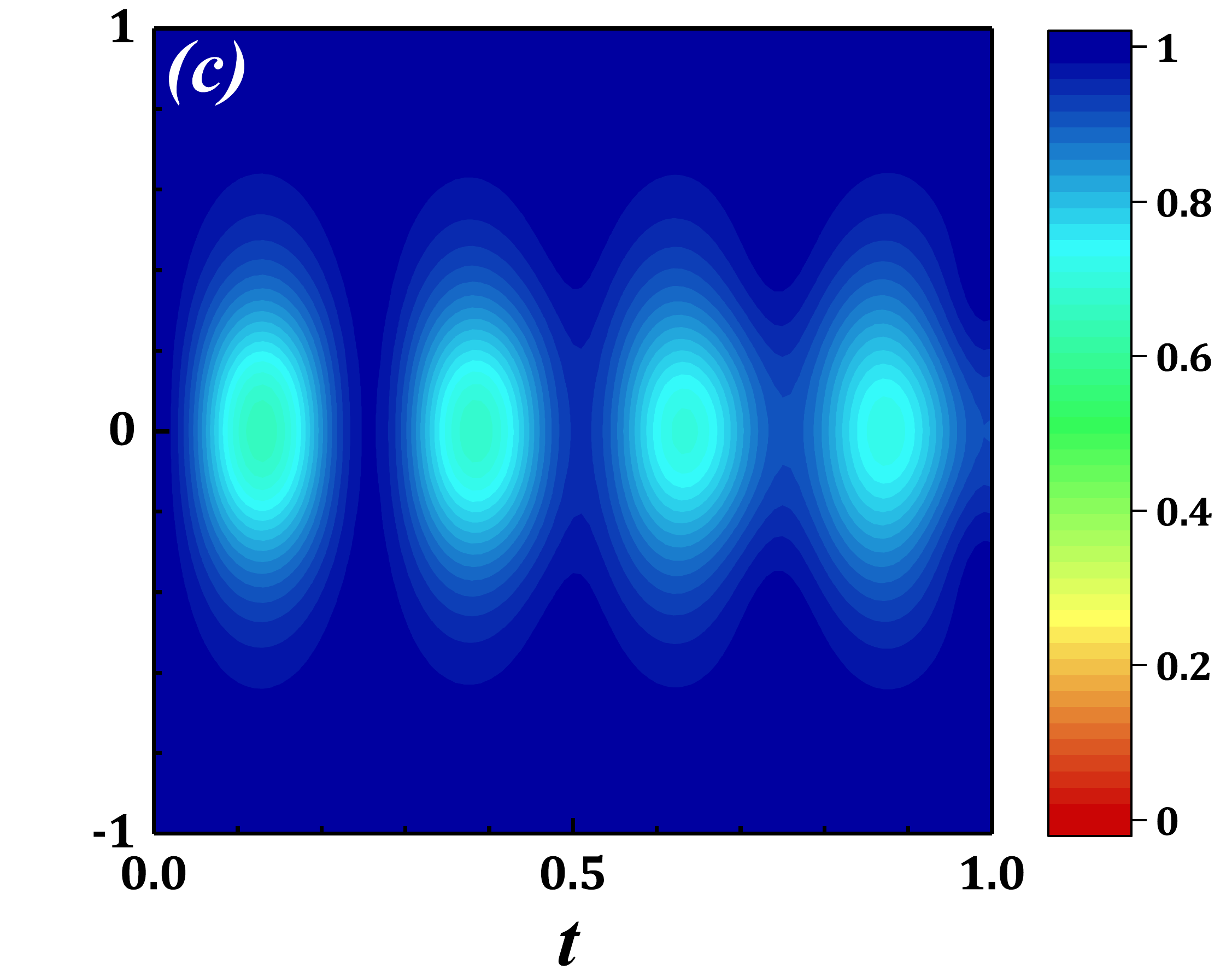}}
\centering
\end{minipage}
\begin{minipage}{\linewidth}
\centerline{\includegraphics[width=0.33\linewidth,height=0.26\linewidth]{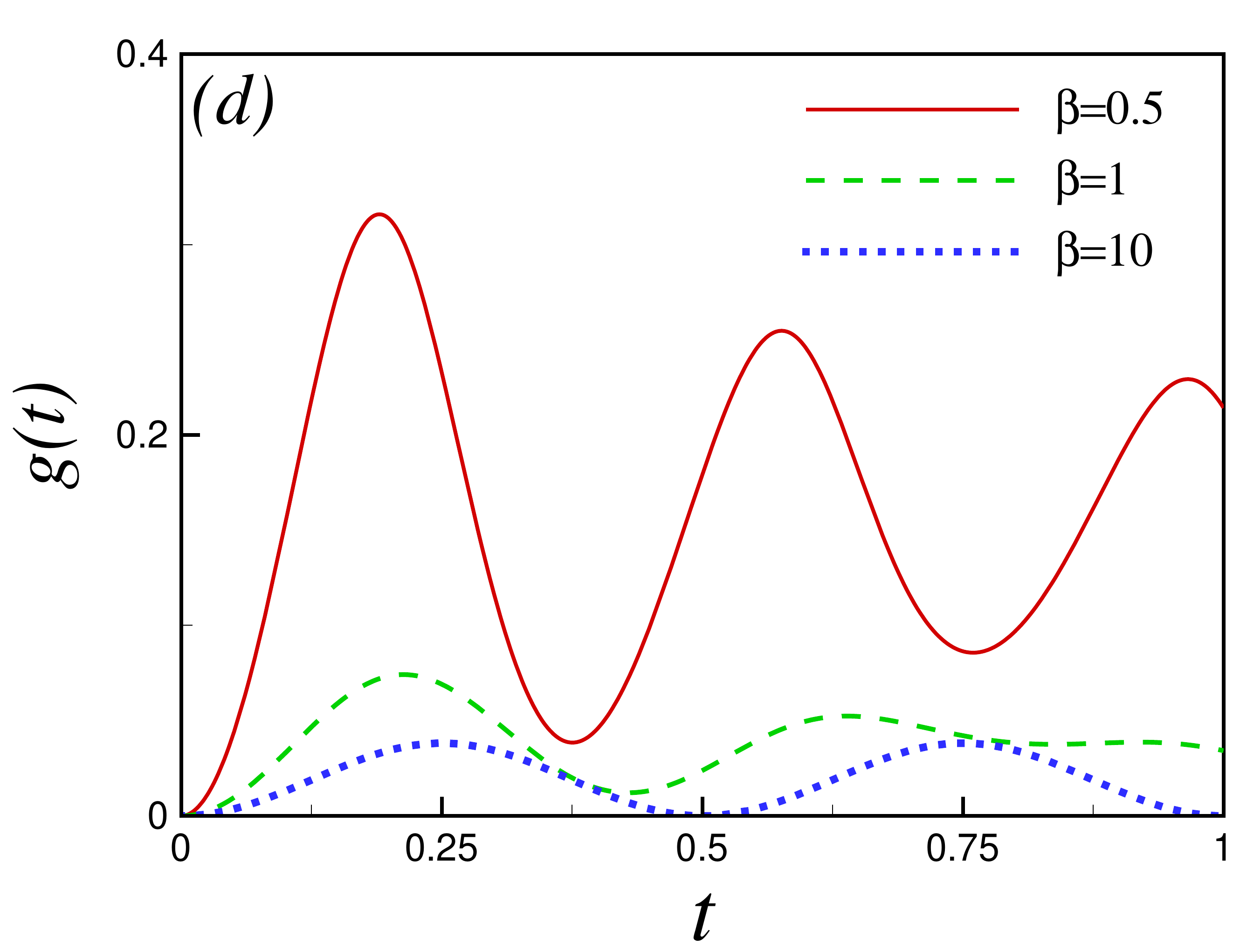}
\includegraphics[width=0.33\linewidth,height=0.26\linewidth]{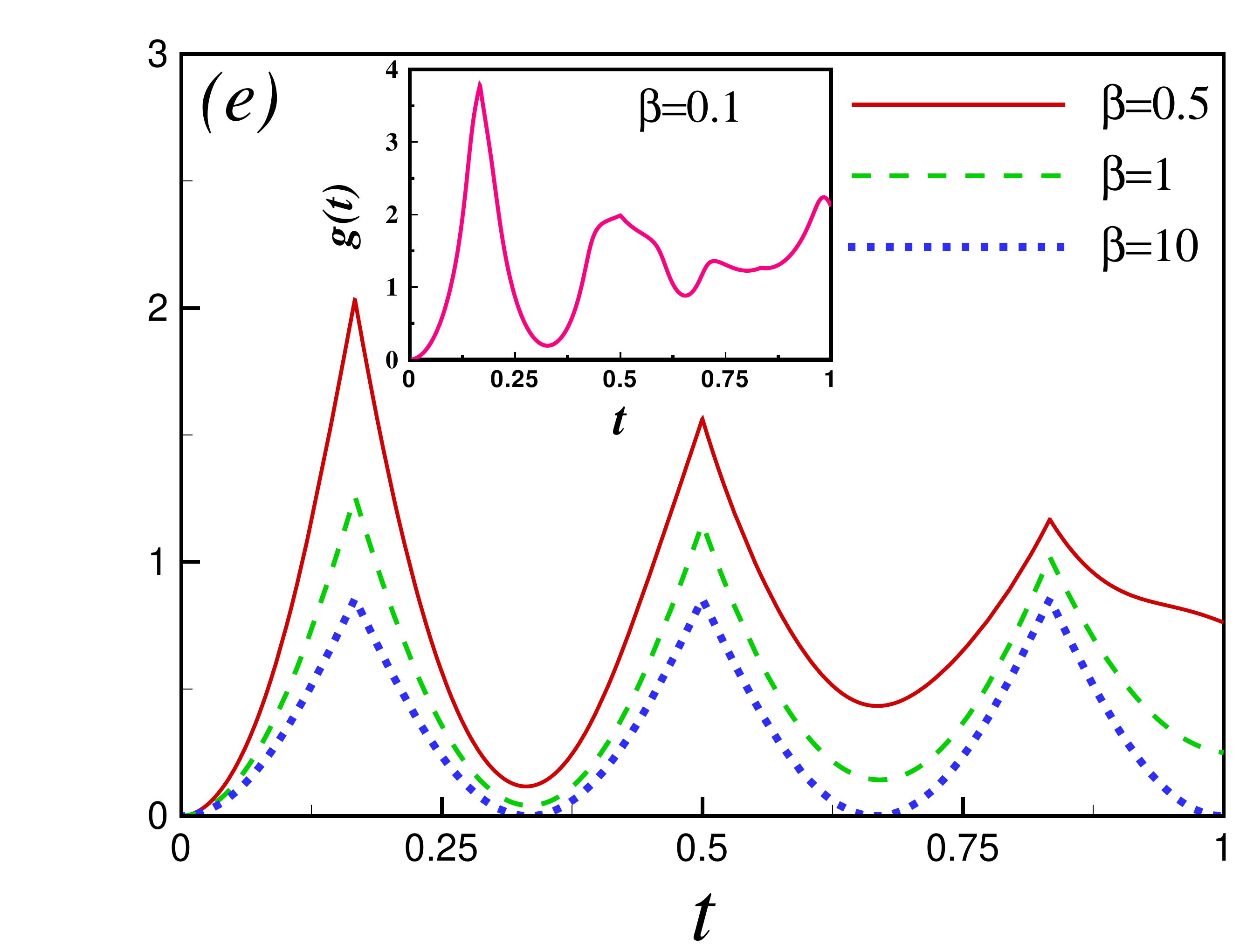}
\includegraphics[width=0.33\linewidth]{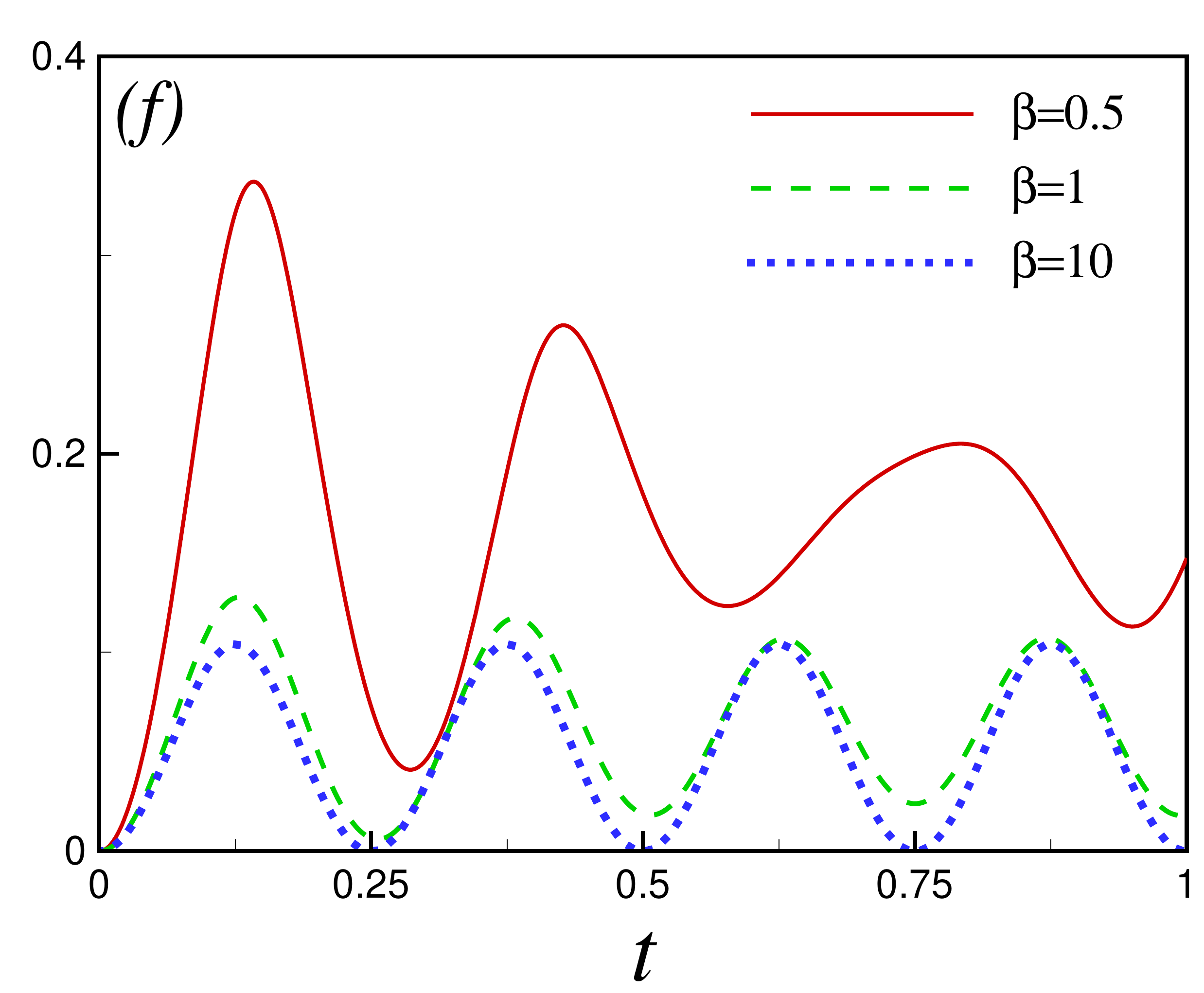}}
\centering
\end{minipage}
\caption{(Color online) The density plot of generalized Loschmidt amplitude versus $t$ and $k$ for $\beta=1$, and (a) $\omega=4\pi$, (b) $\omega=6\pi$ and (c) $\omega=8\pi$.
The rate function of generalized Loschmidt amplitude versus time for different $\beta$
and (d) $\omega=4\pi$, (e) $\omega=6\pi$ and (f) $\omega=8\pi$.
In the whole plots $J_{2}=\pi$, $h_{s}=3\pi$.
}
\label{fig5}
\end{figure*}
%
The sketches of lines of Fisher zeros are shown in Fig. \ref{fig3} for different values of driving frequency. It can be clearly seen that, the imaginary axis is crossed by Fisher zeros lines for the adiabatic regime (Fig. \ref{fig3}(b)).
However, Fisher zeros do not cut the imaginary axis for driving frequencies that drive system nonadiabatically (Figs. \ref{fig3}(a) and \ref{fig3}(c)).

We have also plotted the density plot of time dependent expectation values of quasi-spin components $\langle\sigma^{\alpha}\rangle$, shown in Fig. \ref{fig7} in Appendix \ref{A5}. In the adiabatic regime, it is clearly visible that quasi-spin components oscillate between spin up and down states and cover the range $[-1,1]$.

As stated in Sec.\ref{introduction}, dynamical topological order parameter has been proposed  to represent the topological characteristic associated with DQPTs. The DTOP displays integer values as a function of time and reveals unit magnitude jumps at the critical times at which the
DQPTs occur. The DTOP is given by
%
\begin{eqnarray}
\label{eq17}
\nu_D(t)=\frac{1}{2\pi}\int_0^\pi\frac{\partial\phi^G(k,t)}{\partial k}\mathrm{d}k,
\end{eqnarray}
%
where the geometric phase $\phi^G(k,t)$ is obtained from the total phase $\phi(k,t)$ by subtracting the dynamical
phase $\phi^{D}(k,t)$, i.e., $\phi^G(k,t)=\phi(k,t)-\phi^{D}(k,t)$.
The total phase $\phi(k,t)$ is the phase factor of LA in its polar coordinates representation, i.e. ${\cal L}_{k}(t)=|{\cal L}_{k}(t)|e^{i\phi(k,t)}$, and $\phi^{D}(k,t)=-\int_0^t \langle \psi_{k}^{-}(t')|\mathbb{H}_{k}|\psi_{k}^{-}(t')\rangle dt'$,
are obtained as follows
%
{\small
\begin{eqnarray}
\no
\phi(k,t)=-E^{-}_{k}t &+& \tan^{-1}\Big(\frac{\sin^{2}(\gamma_{k}/2)\sin(\omega t)}{\cos^{2}(\gamma_{k}/2)
+\sin^{2}(\gamma_{k}/2)\cos(\omega t)}\Big),\\
\no
\phi^{D}(k,t)&=&-E^{-}_{k}t+(1-\cos(\gamma))\omega t/2.
\end{eqnarray}
}
%
%
\begin{figure*}
\centerline{\includegraphics[width=0.33\linewidth]{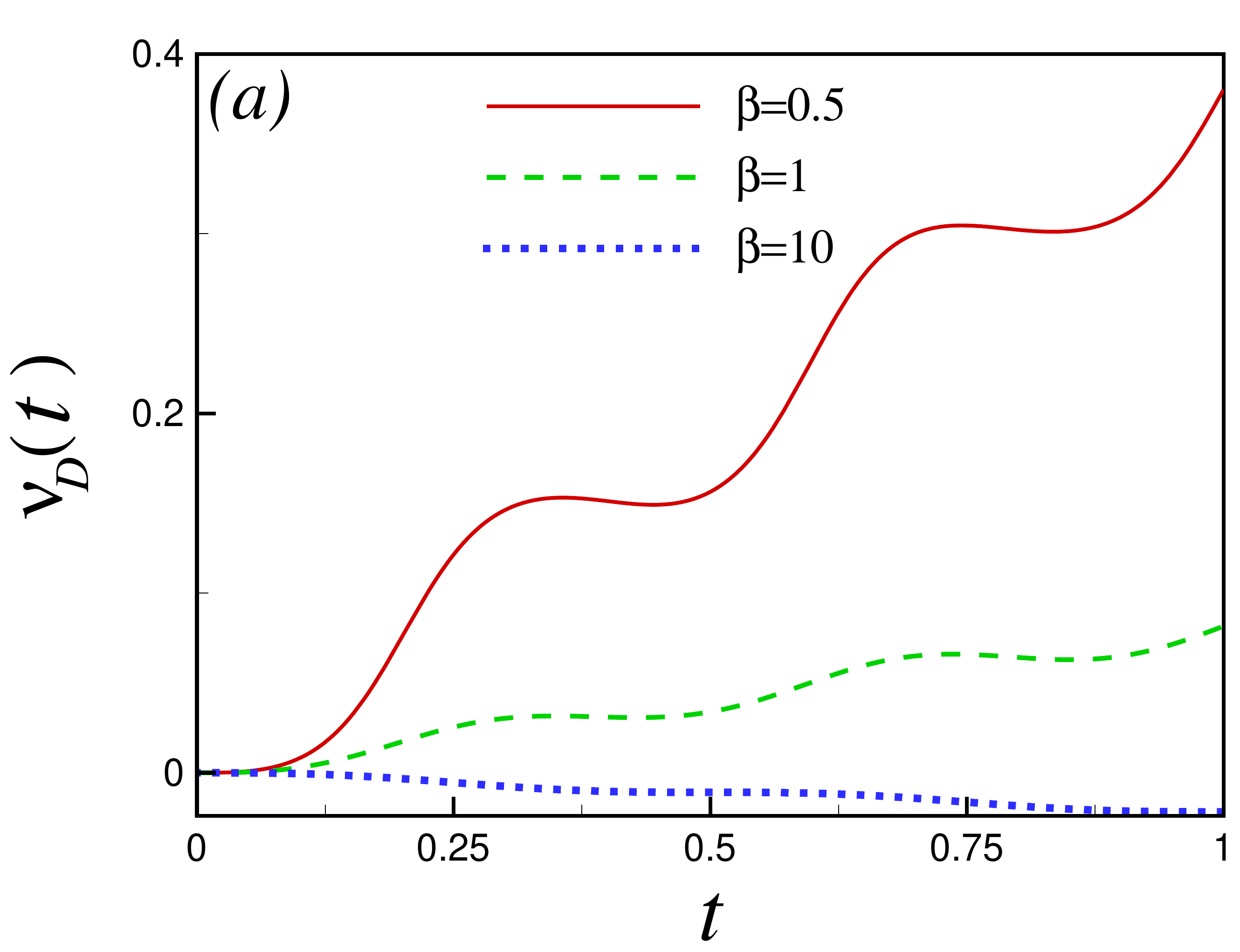}
\includegraphics[width=0.33\linewidth]{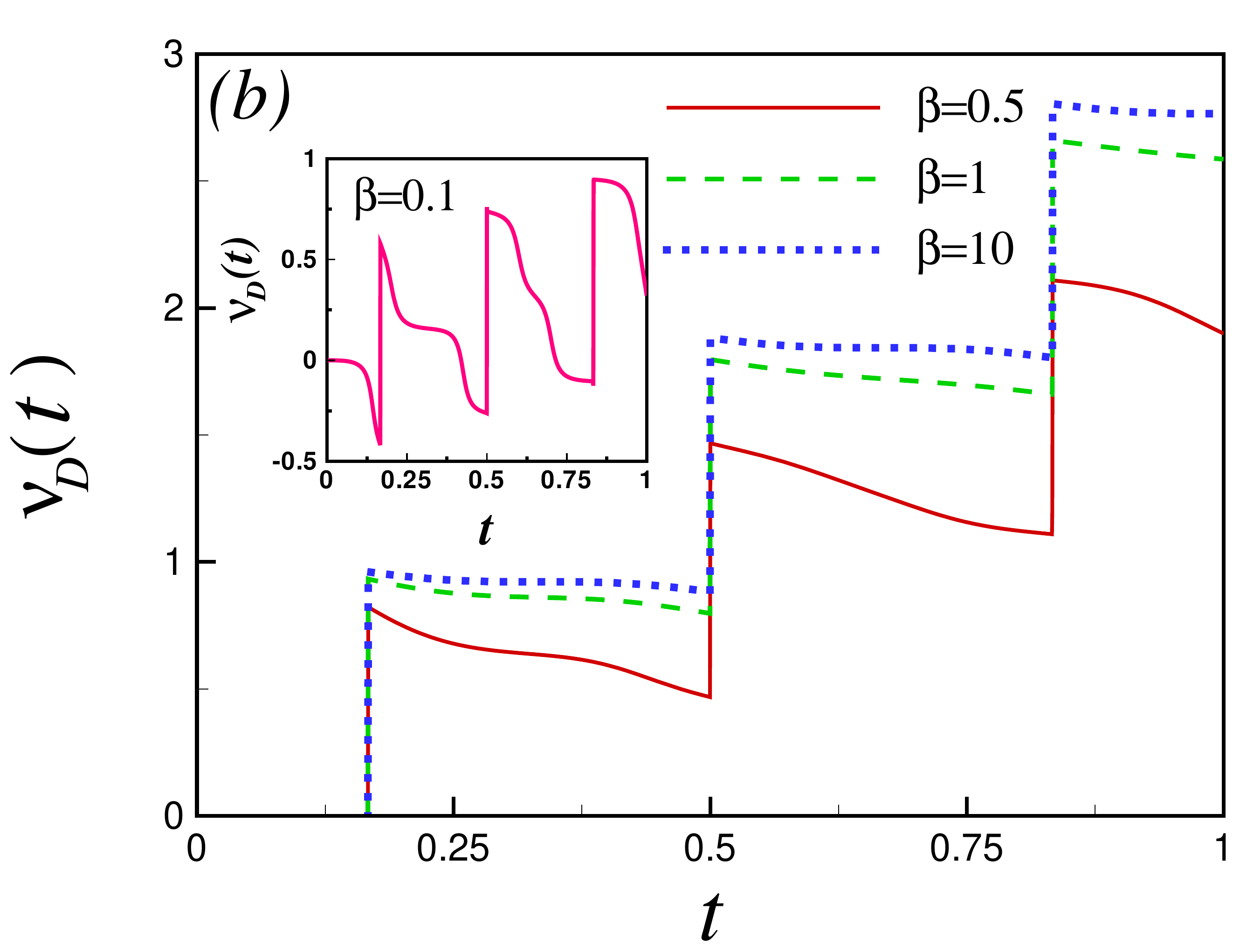}
\includegraphics[width=0.33\linewidth]{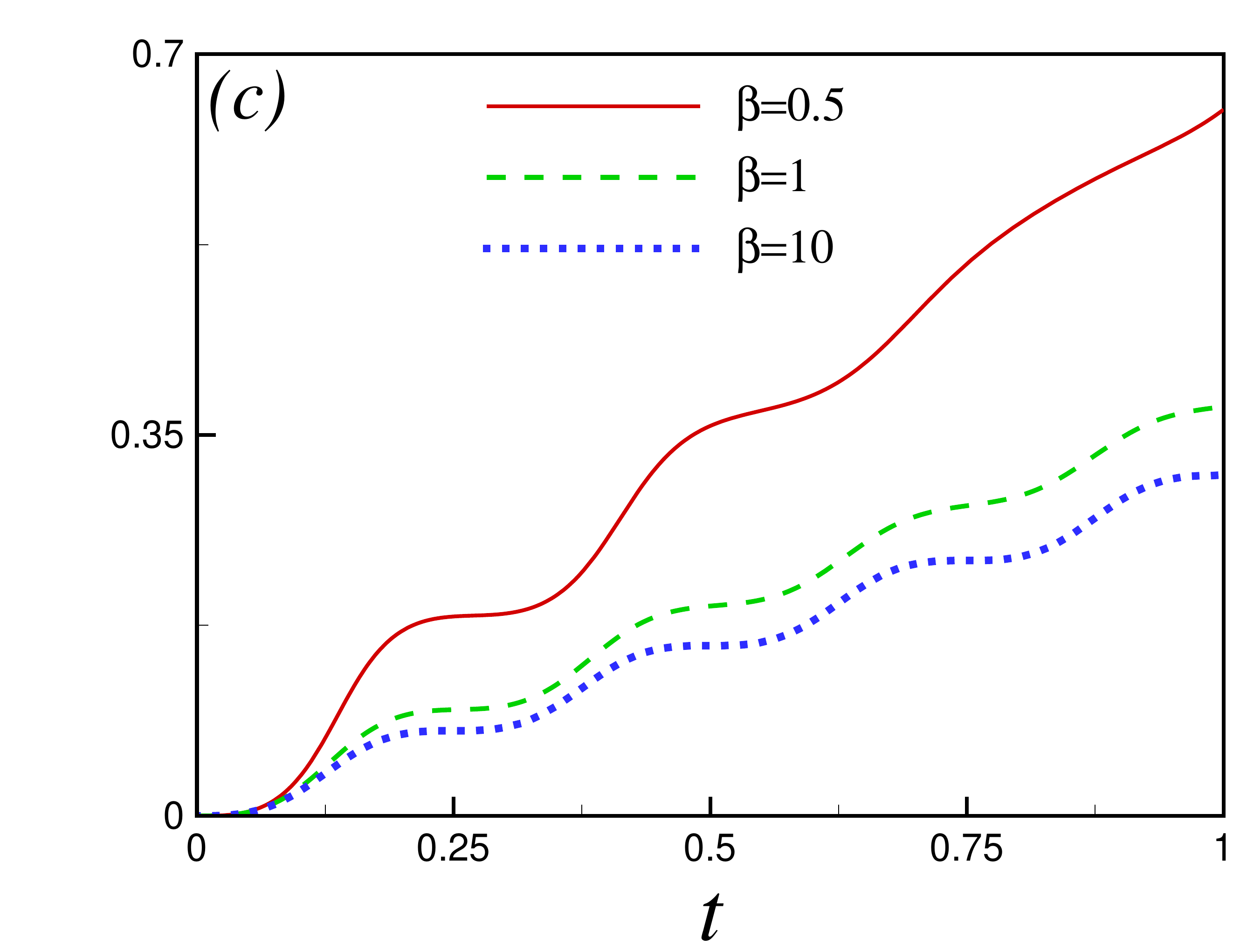}}
\caption{(Color online) The mixed state topological order parameter as a
function of time for different values of $\beta$, $J_{2}=\pi$, $h_{s}=3\pi$, and (a) $\omega=4\pi$, (b) $\omega=6\pi$ and (c) $\omega=8\pi$.}
\label{fig6}
\end{figure*}
%
The DTOP has been displayed in Fig. \ref{fig4} for different values of the driving frequencies.
From Fig. \ref{fig4} one can see that, DTOP smoothly decreases/increases with time in nonadiabatic regime (Figs. \ref{fig4}(a) and \ref{fig4}(c)), which represent the absence of DQPT. The unit jumps in Fig. \ref{fig4}(b) features the topological aspects of DPTs, which happen in the adiabatic regime.

\subsection{Mixed state dynamical phase transition}
In experiments \cite{flaschner2018observation,jurcevic2017direct}, which investigate the far-from-equilibrium theoretical concepts,
the initial state in which system is prepared, is typically not a pure state but rather a mixed state. This leads to propose generalized Loschmidt amplitude (GLA) for mixed thermal states, which perfectly reproduce the nonanalyticities manifested in the pure state DQPTs \cite{bhattacharya2017mixed, heyl2017dynamical}. Here, we investigate mixed state Floquet DQPTs notion in the Floquet Hamiltonian, Eq. (\ref{eq1}). The GLA for thermal mixed state is defined as follows
%
\begin{equation}
{\cal GL}(t) =\prod_{k} {\cal GL}_{k}(t)=\prod_{k}Tr \Big(\rho_{k}(0) U(t)\Big),
\label{eq18}
\end{equation}
%
where $\rho_{k}(0)$ is the mixed state density matrix at time $t=0$, and $U(t)$ is the time-evolution operator. The mixed state density matrix and time-evolution operator of Floquet Hamiltonian (Eq. (\ref{eq1})) is given by
%
\begin{equation}
\label{eq19}
\rho_{k}(0)=\frac{e^{-\beta \mathbb{H}_{k}}}{{\rm Tr}(e^{-\beta \mathbb{H}_{k}})}
=\frac{1}{2}\Big(\mathbb{1}-\tanh(\beta\frac{\Delta_{k}}{2}){{\hat n}_k}\cdot {\vec {\sigma}}\Big),\\
\label{eq20}
U(t)=U_{R}(t)e^{-{\it i}\mathbb{H}_{k}t}=e^{{\it i}\omega(\mathbb{1}-\sigma^{z})t/2}e^{-{\it i}\mathbb{H}_{k}t},
\end{equation}
%
respectively, where $\mathbb{H}_{k}=\frac{1}{2}(\omega\mathbb{1}+\Delta_{k}{\hat n}_k\cdot\vec {\sigma})$,
with ${\hat n}_k=(h_{xy}(k),0,h_{z}(k)-\omega)/\Delta_{k}$ and $\beta=T^{-1}$ is the inverse temperature
with Boltzmann constant $K_{B}=1$.
A rather lengthy calculation yielding an exact expression for GLA
%
\bea
\label{eq21}
{\cal GL}_{k}(t)&=&\frac{1}{\Delta_{k}}\Big[\Delta_{k}\cos(\frac{\omega t}{2})\cos(\frac{\Delta_{k}t}{2})\\
\no
&-&(h_z(k)-\omega)\sin(\frac{\omega t}{2})\sin(\frac{\Delta_{k}t}{2})\\
\no
&+&i\Big(\Delta_{k}\cos(\frac{\omega t}{2})\sin(\frac{\Delta_{k}t}{2})\\
\no
&+&(h_z(k)-\omega)\sin(\frac{\omega t}{2})\cos(\frac{\Delta_{k}t}{2})\Big)\tanh(\beta\Delta_{k}/2)\Big].
\eea
%
The density plot of GLA has been plotted versus time $t$ and $k$ in Figs. \ref{fig5}(a)-(c) for different values of driving frequency at $\beta=1$. As seen, in the adiabatic regime Fig. \ref{fig5}(b), the critical points $k^{\ast}$ and $t^{\ast}$, where GLA becomes zero, are exactly the same as the corresponding one in LA. In the nonadiabatic regime, there is no critical point to get zero for GLA. So we expect that the mixed state DQPT occurs in the adiabatic regime even at finite temperatures. The comparison of Fig. \ref{fig2}(b) with Fig. \ref{fig5}(b) manifests that, GLA shows deformation versus time. Our numerical results show that the deformation enhances by increasing time and temperature.
The rate function of GLA has been plotted versus time in Figs. \ref{fig5}(d)-(f) for different values of $\beta$ and driving frequencies. As is clear, the nonanalyticities in the rate function of GLA appear in the adiabatic regime (Fig. \ref{fig5}(e)). Although GLA correctly reproduces the critical mode $k^{\ast}$, and critical time $t^{\ast}$ observed during the pure state DQPT, the height of cusps decreases by increasing time. However, the height of cusps and its reductional behaviour with time grows by increasing temperature (Fig. \ref{fig5}(e)).
It has to be noted that for temperatures smaller than the temperature associated with the minimum energy gap, the critical modes and times of the mixed state DQPT, remain unaffected. For higher temperatures the finger print of DQPT washed out, which states a cross over to a new regime without DQPT (see inset of Fig. \ref{fig5}(e), which shows the case of $\beta=0.1$).
In Fig. \ref{fig8} (see Appendix \ref{A8}) the time dependent expectation values of quasi-spin components $\langle\sigma^{\alpha}\rangle$, have been plotted at finite temperature. Similar to the pure state case, in the adiabatic regime quasi-spin components oscillate between spin up and down states. It is remarkable to mention that the range, over which quasi-spin components oscillate, shrinks by increasing temperature and at high temperatures their expectation values are approximately constant.

Analogous to the pure state DQPT, topological invariant has been established for mixed state DQPT to display its topological characteristics. In the mixed state DQPT the total phase and dynamical phase are defined as $\phi(k,\beta,t)=Arg\Big[Tr\big(\rho(k,\beta,0)U(t)\big)\Big]$, and $\phi^{D}(k,\beta,t)=-\int_{0}^{t} Tr[\rho(k,\beta,t')H(k,t')]dt'$, respectively. Hence, the topological invariant $\nu_D(t)$ can be calculated for mixed state
using Eq. (\ref{eq17}) in which $\phi^{G}(k,\beta,t)=\phi(k,\beta,t)-\phi^{D}(k,\beta,t)$. After a lengthy calculation, one can obtain  $\phi(k,\beta,t)$ and $\phi^{D}(k,\beta,t)$ for a mixed state (see Appendix \ref{A7}). The mixed state topological invariant has been illustrated in Fig. \ref{fig6} for different values of driving frequencies and $\beta$. One can clearly see that $\nu_D(t)$ exhibits a nearly perfect quantization (unit jump) as a function of time between two successive critical times $t^{\ast}$ in the adiabatic regime, Fig. \ref{fig6}(b).
The quantized structure of $\nu_D(t)$ is only observed as far as temperatures are smaller than the temperature associated with the minimum energy gap. Although abrupt jumps of $\nu_D(t)$ is observed at higher temperatures, it does not show a quantized value to
represent a topological character as seen in the inset of Fig. \ref{fig6}(b), for $\beta=0.1$. This confirms the existence of a crossover temperature below which mixed state DQPT exist and are
signaled by the mixed state DTOP, which are nearly quantized.

\section{Conclusion}
We have studied both pure and mixed states Floquet dynamical quantum phase transition in the periodically driven extended XY model in the presence of staggered magnetic field. We have shown that the proposed Floquet Hamiltonian with $N$ interacting spins can be mapped to $N/2$ noninteracting quasi-spins subjected to the time dependent effective magnetic field (Schwinger-Rabi model). The calculated values of Chern number reveals that there exists topological transition from the nonadiabatic to adiabatic regime. In the adiabatic regime, the quasi-spins follow the time dependent effective magnetic field and then oscillate between up and down states. However, in the nonadiabatic regime the quasi-spins cannot trace the time dependent effective magnetic field and feel an average magnetic field.
We have obtained the range of driving frequency over which
the system experiences adiabatic cyclic process and shows DQPT.
It can be understood that for very high frequencies the adiabatic evolution is lost, while, at very low frequencies the period of cyclic dynamics diverges, which prohibits a recursion to show DQPT at finite time. This justifies the presence of a window frequency
over which the DQPT is observed.
We would like to stress that our model has a single gapless critical point and the observed DQPTs
within a window frequency is different from the observed counterparts in Ref.[\onlinecite{yang2019floquet}],
which possesses two distinct critical points, which define the window frequecy.
We have also obtained the exact expression of Loschmidt amplitude and the generalized Loschmidt amplitude of proposed Floquet system. Our results confirm that both pure and mixed states dynamical phase transitions occur whenever the system evolves adiabatically. In other words, the minimum frequency needed for emerging the dynamical phase transition is equivalent to the threshold driving frequency, which is necessary to drive the system into the adiabatic regime.
Furthermore, we have observed that for the mixed state dynamical phase transition there is a crossover temperature above which the nonanalyticities in the rate function of generalized Loschmidt amplitude and quantization of the mixed state DTOP get completely wiped out.

Floquet systems may show the pre-thermal phase protected by discrete time-translation symmetry
in the absence of disorder and integrals of motion \cite{Prosen1998,Luca2014,Lazarides2014,Ponte2015,Mori2016,Khemani2016,Abanin2017,Else2017}. It would
be an exciting topic to investigate the presence of DQPT in a pre-thermal regime, which can be
realized by adding interaction/perturbation to the model we discussed.

\section{Acknowledgements}
A. L. would like to acknowledge the support from Sharif University
of Technology under Grant No. G960208.


\appendix

\section{Spinless fermion transformation\label{A1}}
The Hamiltonian, Eq. (\ref{eq1}), can be diagonalized by means of the Jordan-Wigner transformation
%
\begin{eqnarray}
&&S^{+}_{n}= S^{x}_{n} + {\it i}S^{y}_{n}=\prod_{m=1}^{n-1}(1-2c_{m}^{\dagger}c_{m})c_{n}^{\dagger},
\nonumber\\
&&S^{-}_{n}= S^{x}_{n} - {\it i}S^{y}_{n} = \prod_{m=1}^{n-1}c_{n}(1-2c_{m}^{\dagger}c_{m}),
\nonumber\\
&&S^{z}_{n} = c_{n}^{\dagger}c_{n} -\frac{1}{2}
\label{eq22},
\end{eqnarray}
%
which transforms spins into fermion operators $c_{n}$, and $c^{\dagger}_{n}$. The essential step here is to define independent fermions at site $n$, $ c_{n-1/2}^{A} \equiv c_{2n-1}, \qquad c_{n}^{B} \equiv c_{2n} $. This can be regarded as dividing the chain into diatomic elementary cells. Introducing the Nambu spinor $\Gamma^{\dagger}_k=(c_{k}^{\dagger B},~c_{k}^{\dagger A})$ the Fourier transformed Hamiltonian can be expressed as sum of independent terms (Eq. (\ref{eq2})) given by
%
{\small
\begin{eqnarray}
\label{eq23}
&& H_{k}(t)=\frac{J_{1}}{2} \Big[2e^{i\omega t} \cos(k/2) c_k^{A\dagger} c_k^B + 2e^{-i\omega t} \cos(k/2) c_k^{B\dagger} c_k^A
\no \\
&-&(J_2 \cos(k)+ 2h_s) c_k^{A\dagger} c_k^A + (J_2 \cos(k)+ 2h_s) c_k^{B\dagger} c_k^B \Big],
\end{eqnarray}
}
%
where the wave number is $k=(2p-1)\pi/N$, $p=-N/4+1,-N/4+2, \cdots  N/4$.

\section{Time-independent Hamiltonian\label{A2}}

By solving the time-dependent Schr\"{o}dinger equation, we can obtain the eigenvalues and eigenvectors of Hamiltonian $H_{k}(t)$
%
\begin{equation}
\label{eq24}
i\frac{d}{dt}|\psi^{\pm}_{k}(t)\rangle=H_{k}(t)|\psi^{\pm}_{k}(t)\rangle.
\end{equation}
%
The exact solution of time dependent Schr\"{o}dinger equation is found by going to the rotating frame given by the periodic unitary transformation
%
\begin{equation}
\label{eq25}
U_{R}(t)=e^{i\omega(\mathbb{1}-\sigma^{z})t/2}=\left(
\begin{array}{cc}
1 & 0 \\
0 & e^{i\omega t} \\
\end{array}
\right).
\end{equation}
%
In the rotating frame an eigenstate is given by $|\chi^{\pm}_{k}\rangle=U_{R}^{\dagger}(t)|\psi^{\pm}_{k}(t)\rangle$.
Substituting the transformed eigenstate into the time dependent Schr\"{o}dinger equation, we arrive at the following time independent
Hamiltonian in the rotating frame approach
%
\begin{equation}
\label{eq26}
\mathbb{H}_{k}=U_{R}^{\dagger}(t)H_{k}(t)U_{R}(t)-iU_{R}^{\dagger}(t)\frac{dU_{R}(t)}{dt}.
\end{equation}
%
So the eigenvalues and eigenstate of the original Hamiltonian can be obtained by solving a time independent
Schr\"{o}dinger equation
%
\begin{equation}
\label{eq27}
\mathbb{H}_{k}|\chi^{\pm}_{k}\rangle=E^{\pm}_{k}|\chi^{\pm}_{k}\rangle.
\end{equation}
%

\section{Floquet system\label{A3}}
Adiabatic and nonadiabatic evolution can be understood in terms of different driving frequencies
within the Floquet theory. A Floquet can be expressed by the following
eigenvalue problem ($\nu=\pm$):
%
\begin{equation}
\label{eq28}
H^{F}_{k}(t)|u^{\nu}_{k}(t)\rangle=\varepsilon^{\nu}_{k}|u^{\nu}_{k}(t)\rangle,
\end{equation}
%
where, $H^{F}_{k}(t)=H_{k}(t)-i\frac{d}{dt}$ and $|u^{\nu}_{k}(t)\rangle$, are the Floquet operator and the Floquet state, respectively, where $\varepsilon^{\nu}_{k}$, is the quasi-energy.
Hence, Eq. (\ref{eq28}) can be rewritten in the form of time dependent Schr\"{o}dinger equation
$[H_{k}(t)-\varepsilon^{\nu}_{k}]|u^{\nu}_{k}(t)\rangle=i\frac{d}{dt}|u^{\nu}_{k}(t)\rangle$.
Within this formulation, a periodically time dependent Hamiltonian has a complete set of solutions $|\psi^{\nu}_{k}\rangle$ that are separable into a product of a phase factor $e^{-i\varepsilon^{\nu}_{k}t}$ and a time-periodic Floquet state $|u^{\nu}_{k}(t)\rangle$, i.e.,
%
\begin{equation}
\no
|\psi^{\nu}_{k}(t)\rangle=e^{-i\varepsilon^{\nu}_{k}t}|u^{\nu}_{k}(t)\rangle.
\end{equation}
%
 Similar to
the exact solution in section \ref{schrodingerEQ}, the above time-dependent Hamiltonian is transformed to the corresponding time-independent form:
%
\begin{equation}
\mathbb{H}^{F}_{k}|\chi^{\nu}_{k}\rangle=\epsilon^{\nu}_{k}|\chi^{\nu}_{k}\rangle
\label{eq29},
\end{equation}
%
where, the time independent Floquet operator $\mathbb{H}^{F}_{k}$ is in the rotating frame expressed as
$\mathbb{H}^{F}_{k}=\mathbb{H}_{k}-\varepsilon^{\nu}_{k}\mathbb{1}$. The eigenstates $|\chi^{\nu}_{k}\rangle$ are given by Eq. (\ref{eq5}) and their corresponding eigenvalues $\epsilon^{\nu}_{k}$ are $E^{\nu}_{k}-\varepsilon^{\nu}_{k}$, in which $E^{\nu}_{k}$ is expressed in Eq. (\ref{eq4}). Then the time-evolved state in the rotating frame is obtained as
%
\begin{eqnarray}
\label{eq30}
|\chi^{\nu}_{k}(t)\rangle=e^{-i(E^{\nu}_{k}-\varepsilon^{\nu}_{k})t}|\chi^{\nu}_{k}\rangle.
\end{eqnarray}
%
Consequently the Floquet state is calculated as
%
\begin{equation}
\label{eq31}
|u^{\nu}_{k}(t)\rangle=e^{i\omega t(\mathbb{1}-\sigma^{z})/2} e^{-i(E^{\nu}_{k}-\varepsilon^{\nu}_{k})t}|\chi^{\nu}_{k}\rangle.
\end{equation}
%
%
\begin{figure*}[t]
\begin{minipage}{\linewidth}
\centerline{\includegraphics[width=0.33\linewidth]{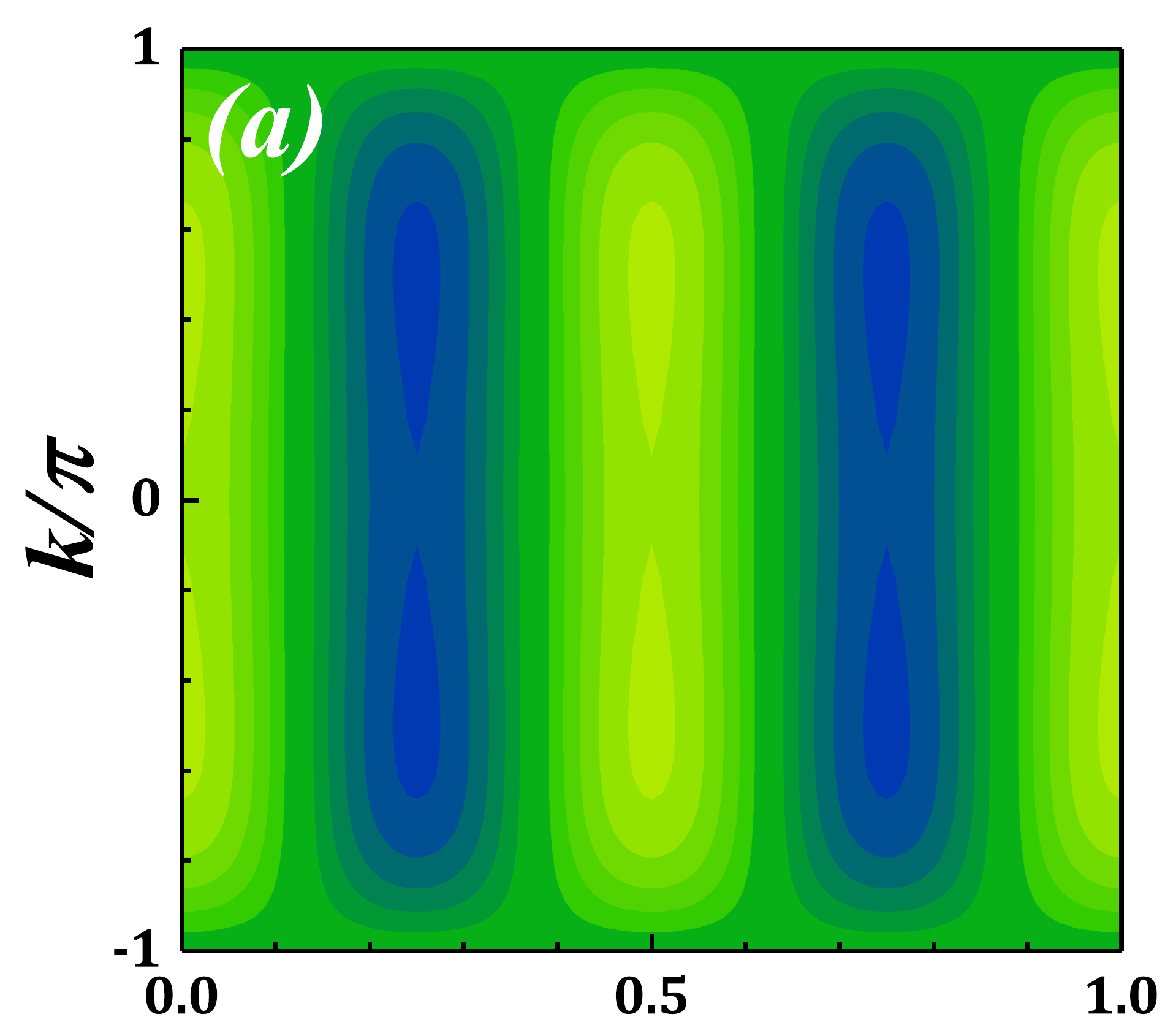}
\includegraphics[width=0.30\linewidth]{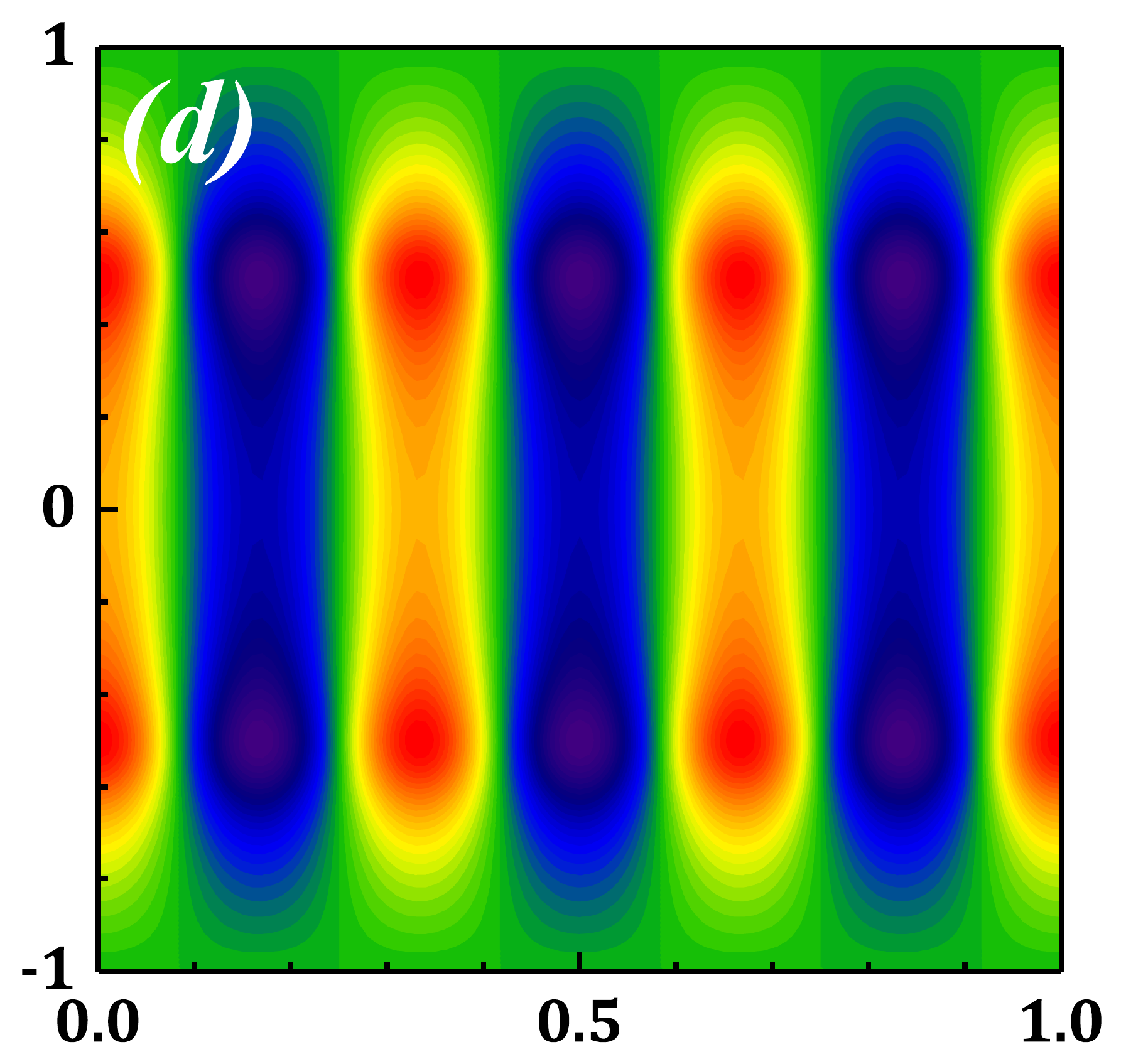}
\includegraphics[width=0.36\linewidth]{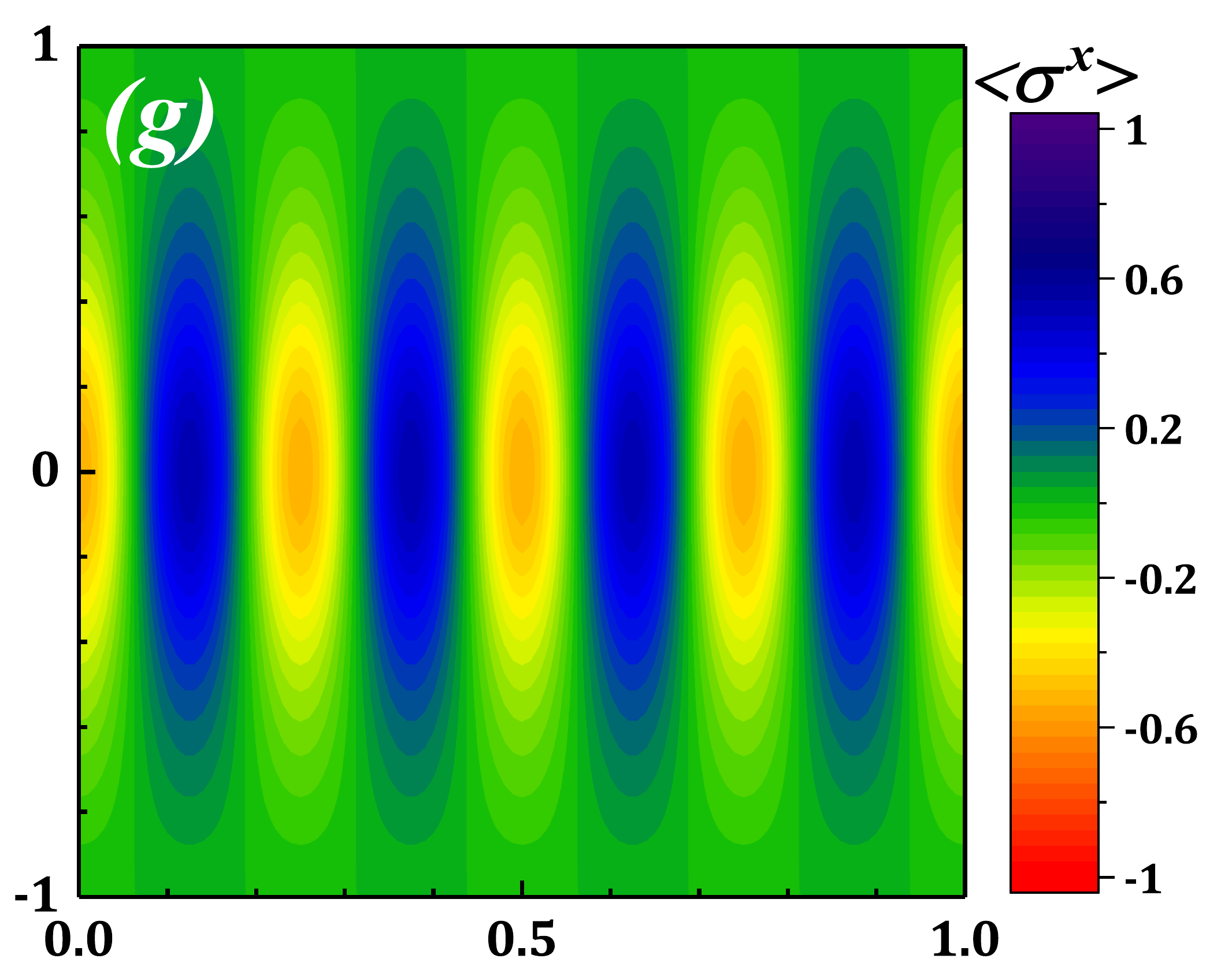}}
\centering
\end{minipage}
\begin{minipage}{\linewidth}
\centerline{\includegraphics[width=0.33\linewidth]{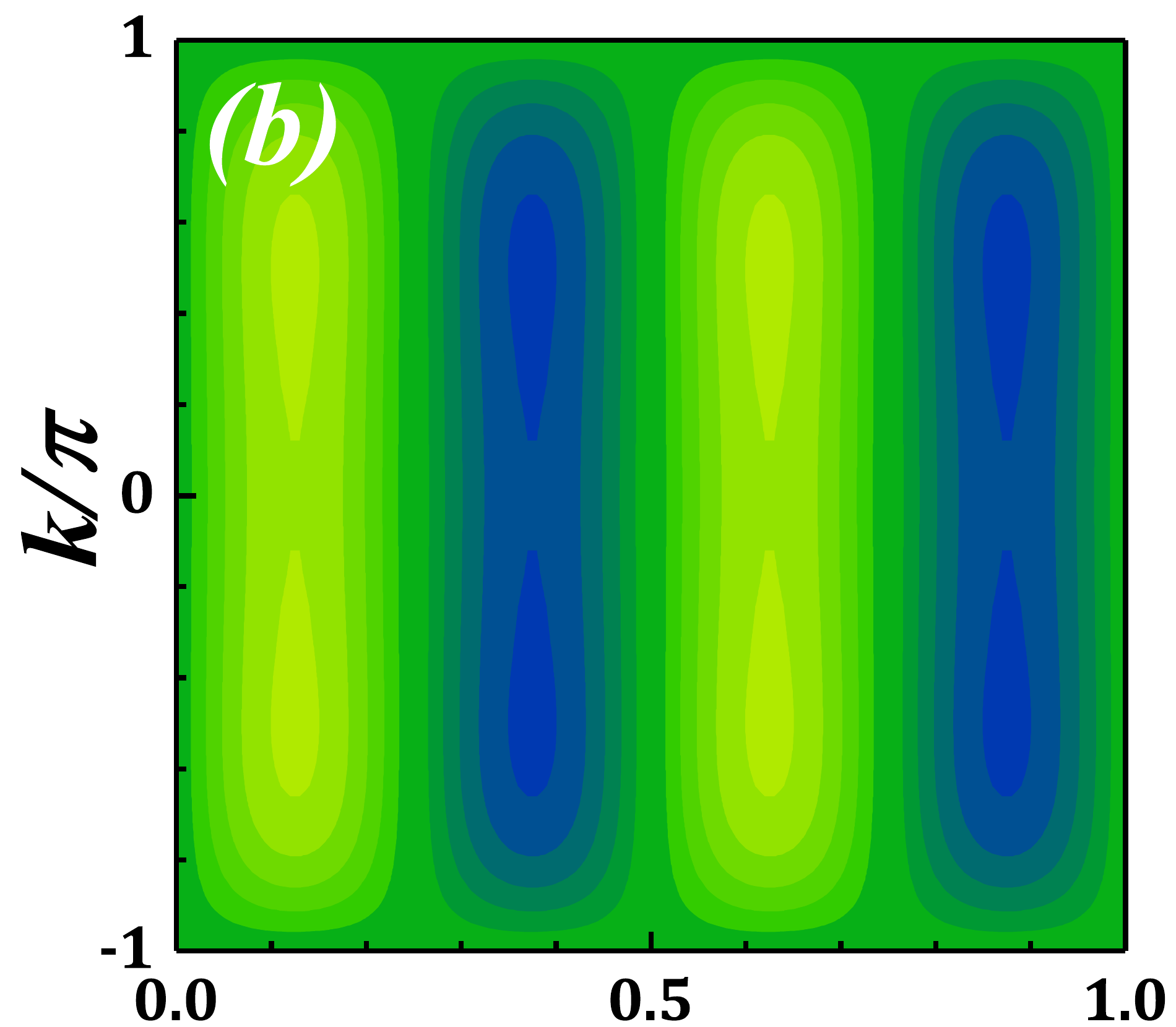}
\includegraphics[width=0.30\linewidth]{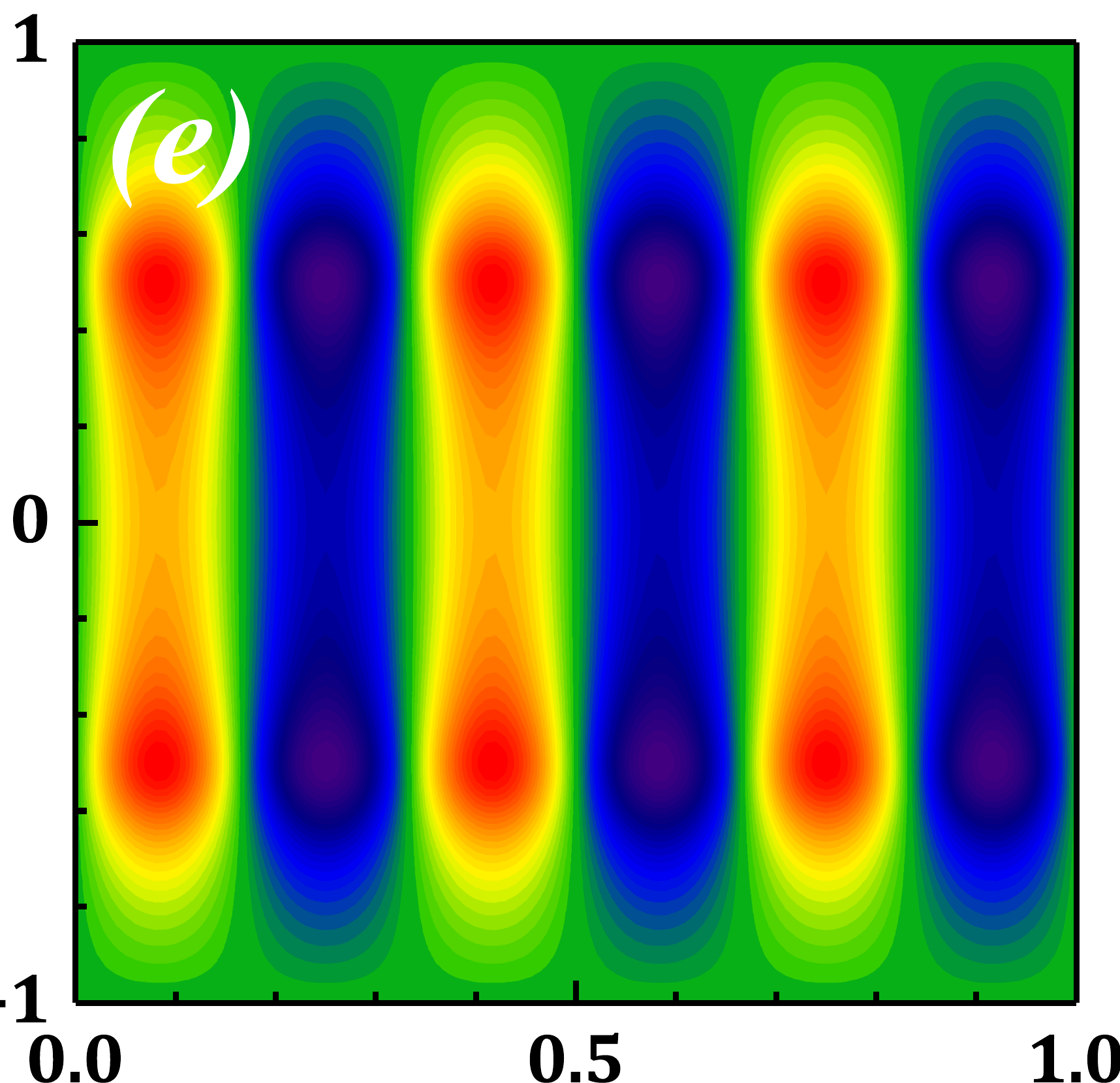}
\includegraphics[width=0.365\linewidth]{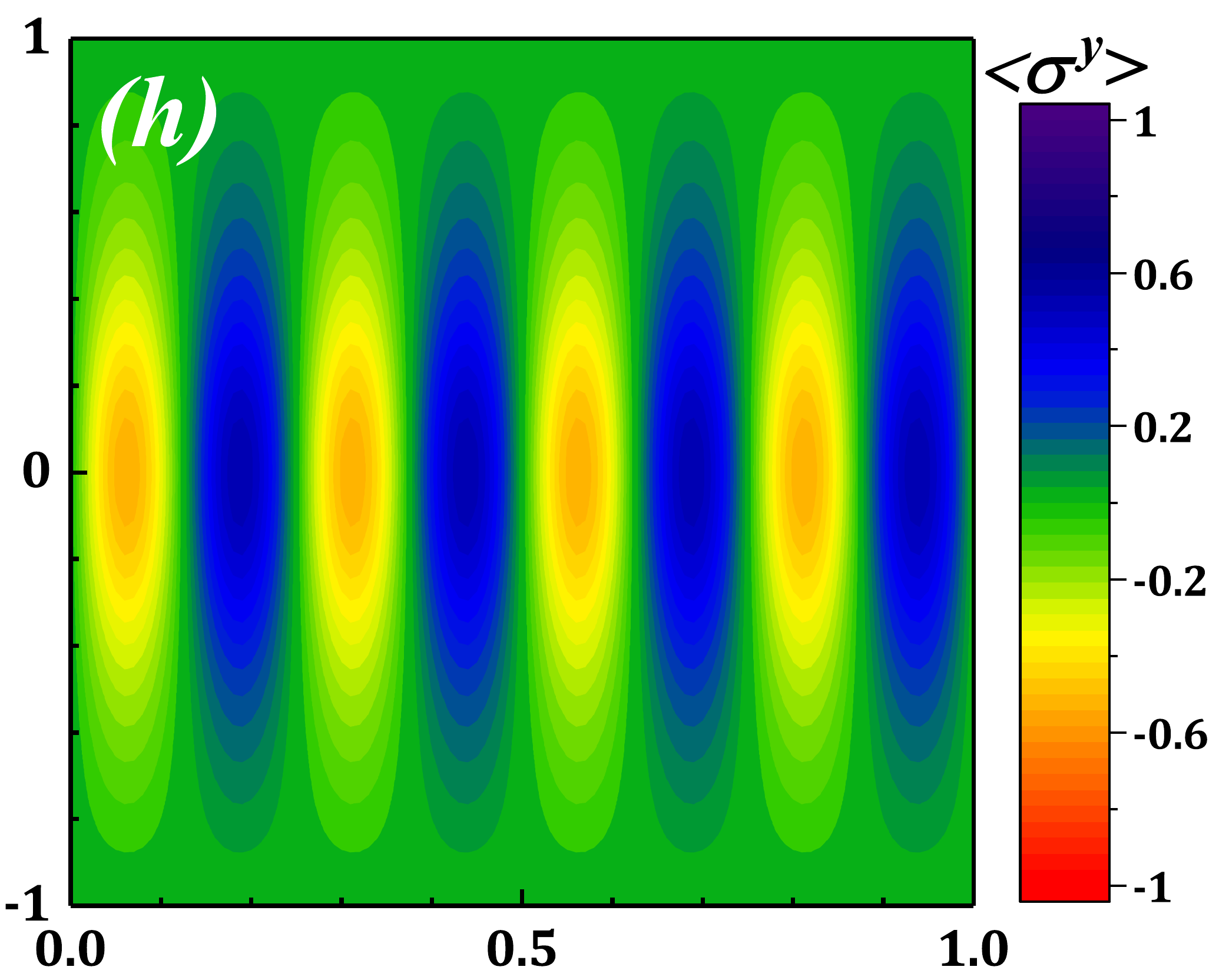}}
\centering
\end{minipage}
\begin{minipage}{\linewidth}
\centerline{\includegraphics[width=0.334\linewidth]{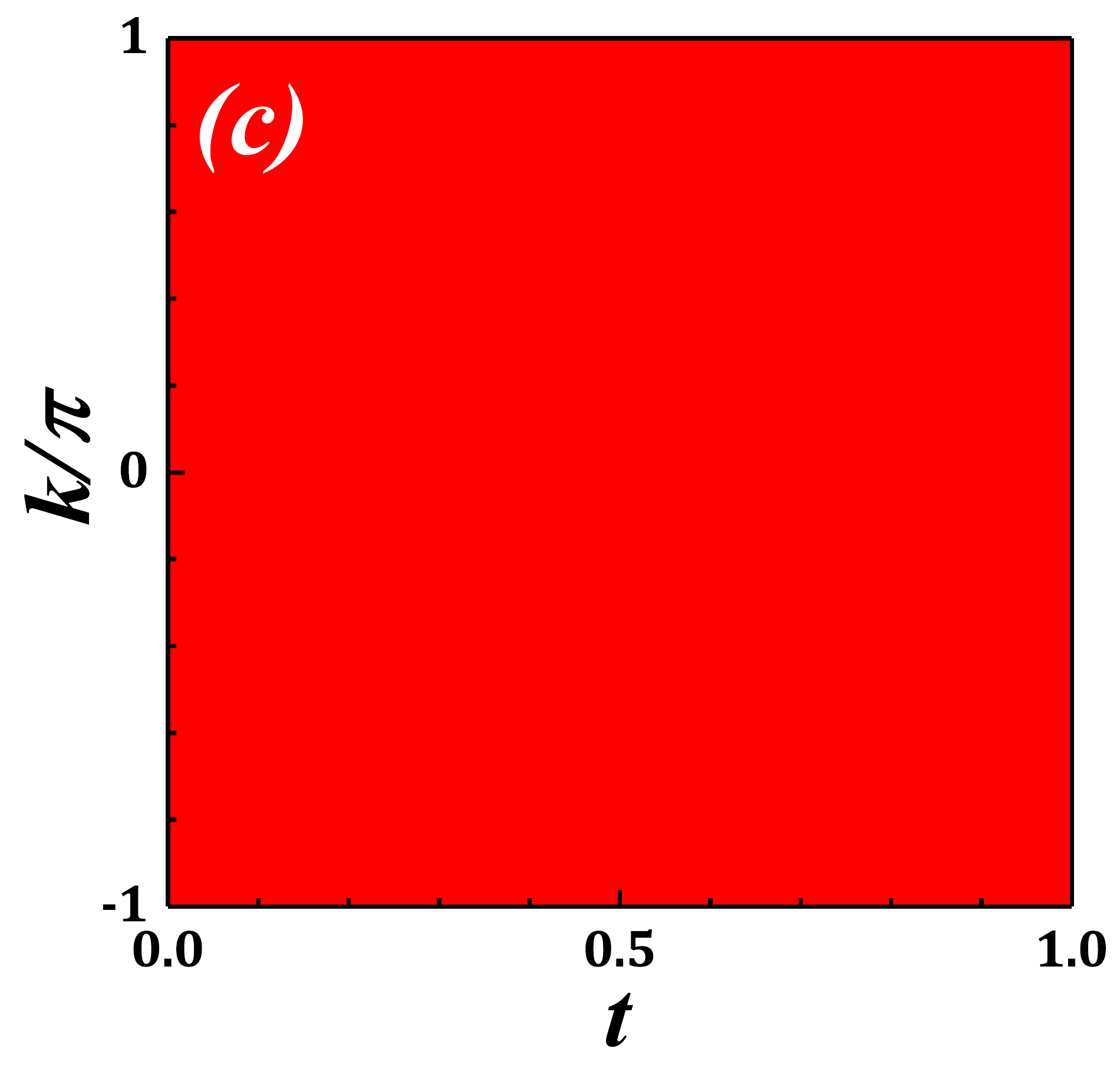}
\includegraphics[width=0.302\linewidth]{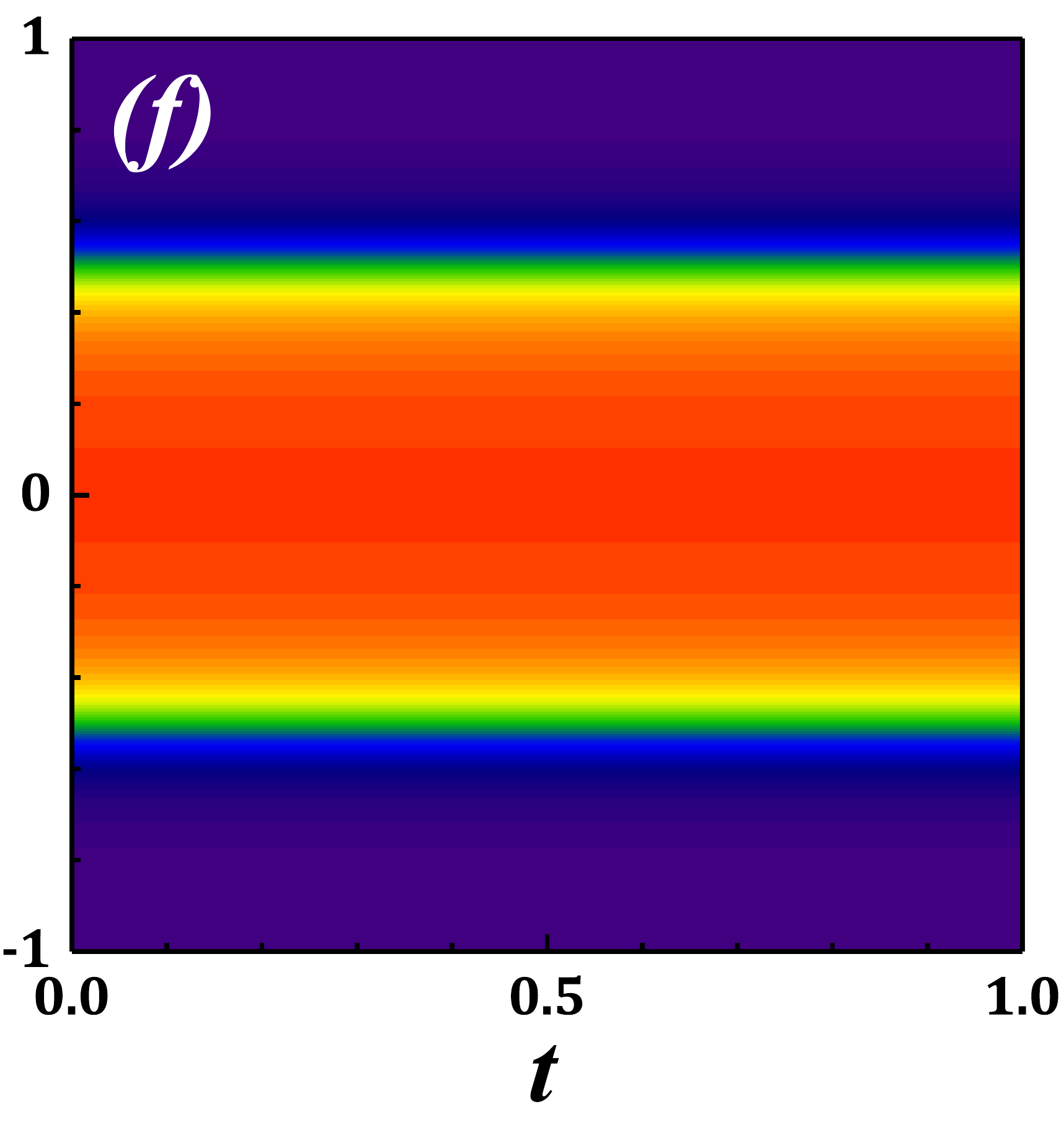}
\includegraphics[width=0.364\linewidth]{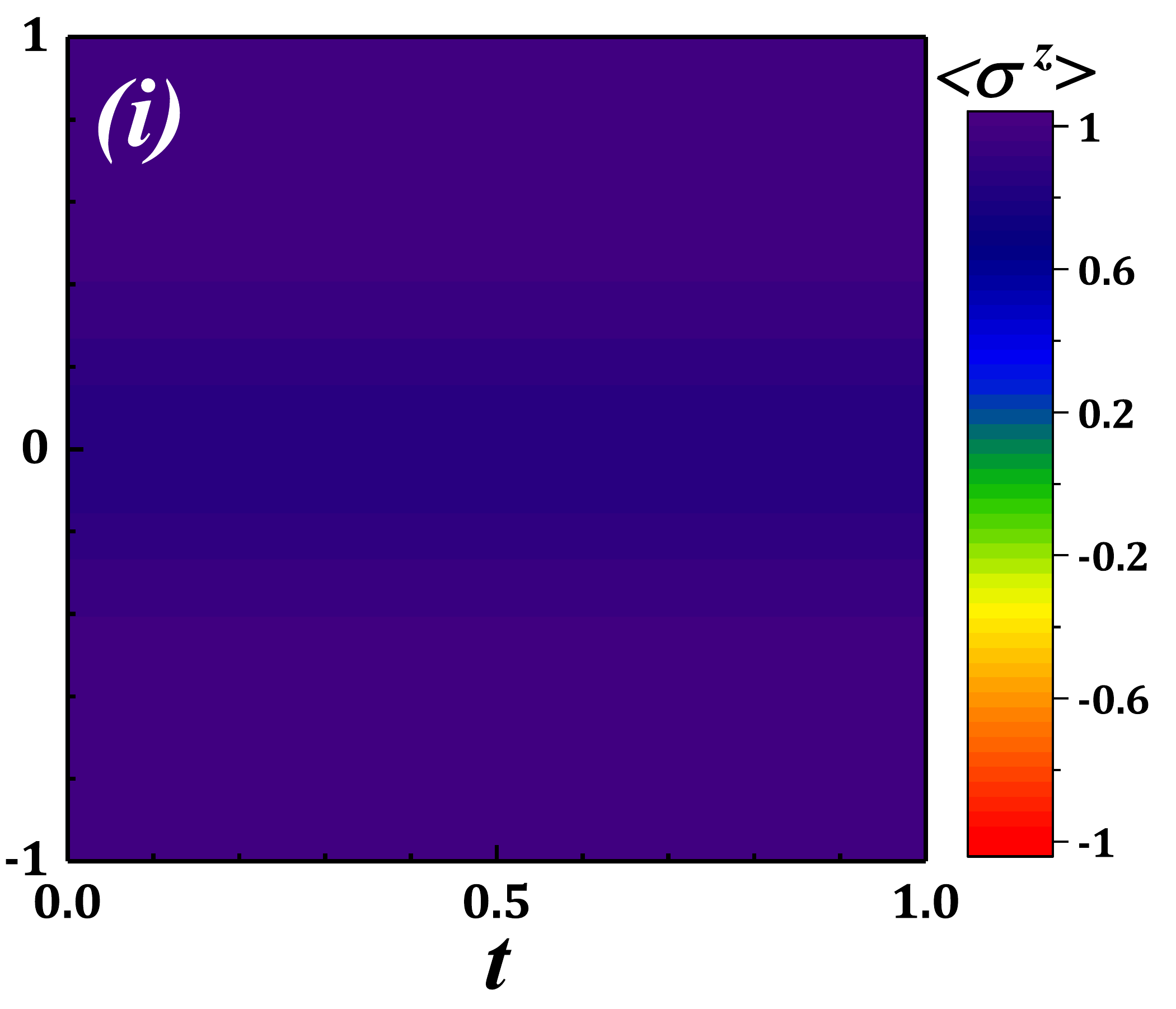}}
\centering
\end{minipage}
\caption{(Color online) Density plot of the expectation values of
$\sigma^{x}$, $\sigma^{y}$, and $\sigma^{z}$ for the pure state dynamical phase
transition versus $t$ and $k$ for $J_{2}=\pi$, $h_{s}=3\pi$, and
(a-c) $\omega=4\pi$, (d-f) $\omega=6\pi$
and (g-i) $\omega=8\pi$.}
\label{fig7}
\end{figure*}
%
Accordingly, Floquet states require time periodicity, which fixes the values of quasi-energies to
%
\begin{equation}
\label{eq32}
\varepsilon^{\nu}_{k}=E^{\nu}_{k}-m\omega, ~~m \in \mathbb{Z}.
\end{equation}
%
Then, the time-dependent Floquet states and time-dependent eigenstates of Floquet Hamiltonian of Eq. (\ref{eq1}) is obtained as
%
\begin{equation}
\label{eq33}
|u^{\nu}_{k}(t)\rangle=e^{i\omega t(\mathbb{1}-\sigma^{z})/2} e^{-i m\omega t}|\chi^{\nu}_{k}\rangle,\\
\no
|\psi^{\nu}_{k}(t)\rangle=e^{-i\varepsilon_{k}^{\nu} t} |u^{\nu}_{k}(t)\rangle.
\end{equation}
%

\section{Quasi-energy and geometric phase\label{A4}}
The Floquet states differ from the exact eigenstates of Schr\"{o}dinger equation, only in a phase factor, which facilitate our calculations to study the topological quantities. Hence, the Berry connection $A$ is obtained as:
%
\begin{equation}
\label{eq34}
A^{\nu}_{k}=\langle u^{\nu}_{k}(t)|id|u^{\nu}_{k}(t)\rangle=A^{\nu}_{k}(t)dt+A^{\nu}_{k}(\theta)d\theta,
\end{equation}
%
where,  $A^{\nu}_{k}(t)$ and $A^{\nu}_{k}(\theta)$ are given by:
%
\begin{eqnarray}
\label{eq35}
A^{\nu}_{k}(t)&=&\langle u^{\nu}_{k}(t)|i\partial_{t}|u^{\nu}_{k}(t)\rangle\\
\no
&=&\frac{\omega}{2}[\langle u^{\nu}_{k}(t)|\sigma^{z}|u^{\nu}_{k}(t)\rangle+(2m-1)]\\
\no
&=&\frac{\omega}{2}[-\nu \cos(\gamma)+(2m-1)],\\
\no
A^{\nu}_{k}(\theta)&=&\langle u^{\nu}_{k}(t)|i\partial_{\theta}|u^{\nu}_{k}(t)\rangle.
\end{eqnarray}
%
%
\begin{figure*}[t]
\begin{minipage}{\linewidth}
\centerline{\includegraphics[width=0.33\linewidth]{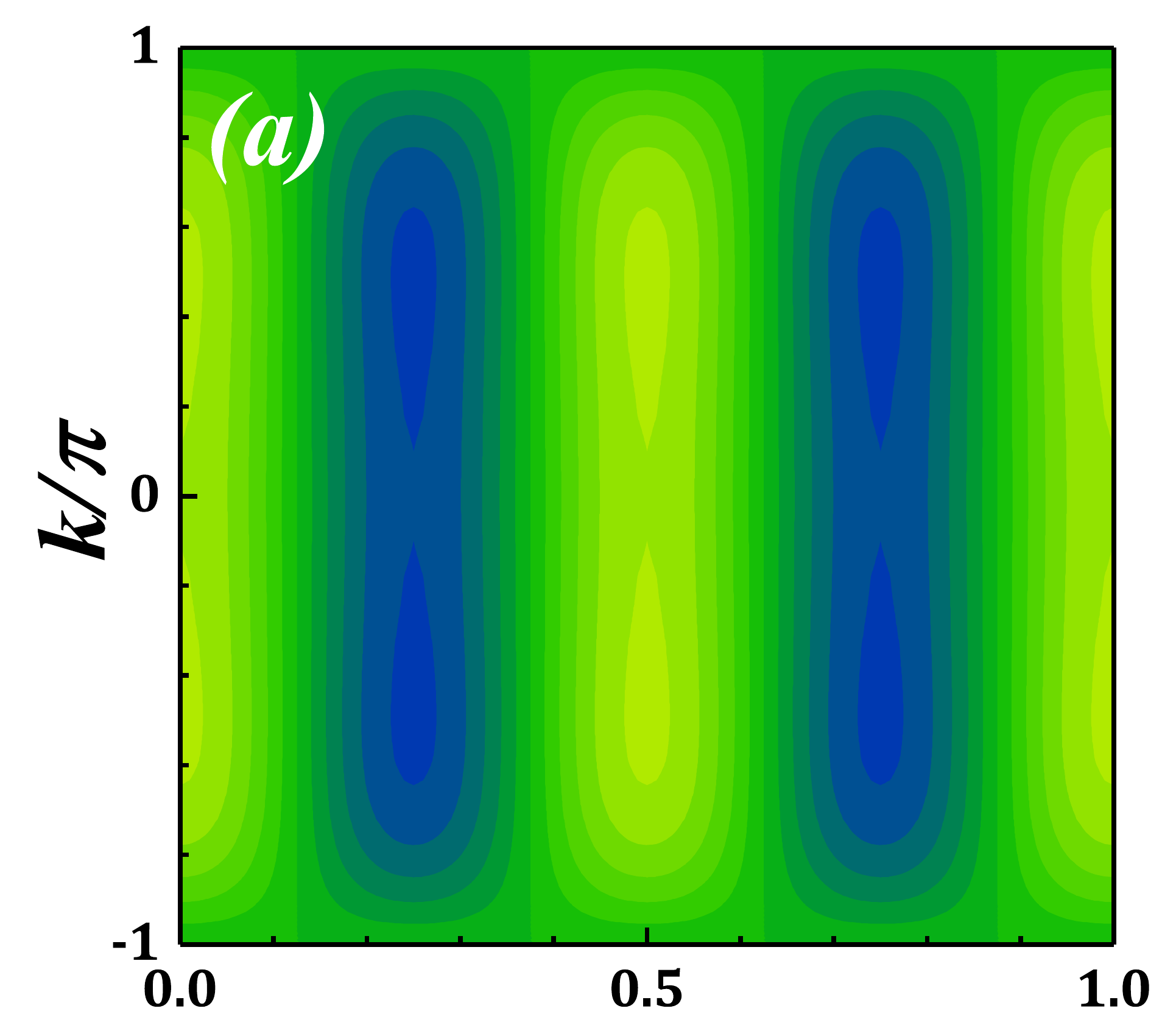}
\includegraphics[width=0.305\linewidth]{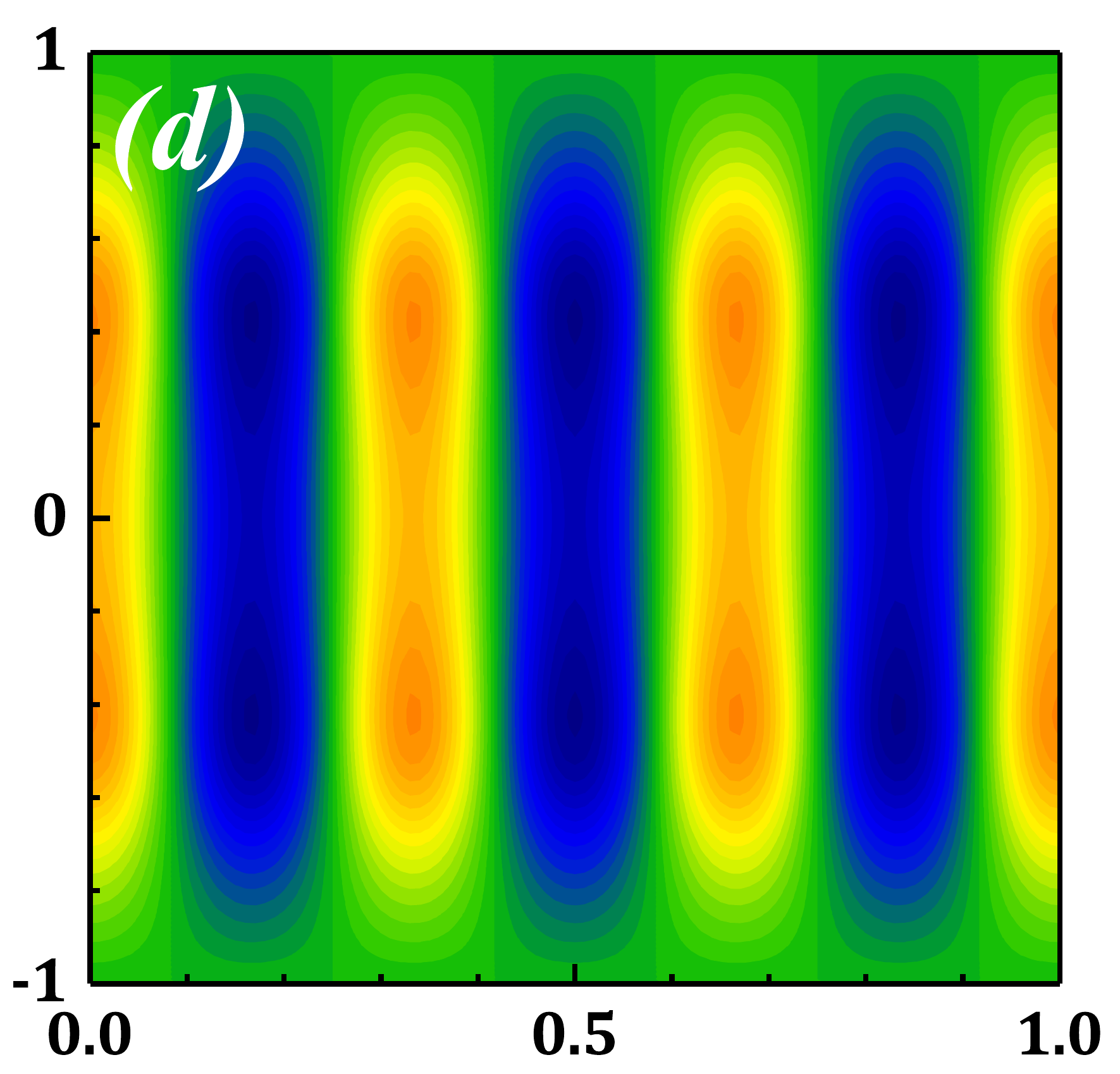}
\includegraphics[width=0.365\linewidth]{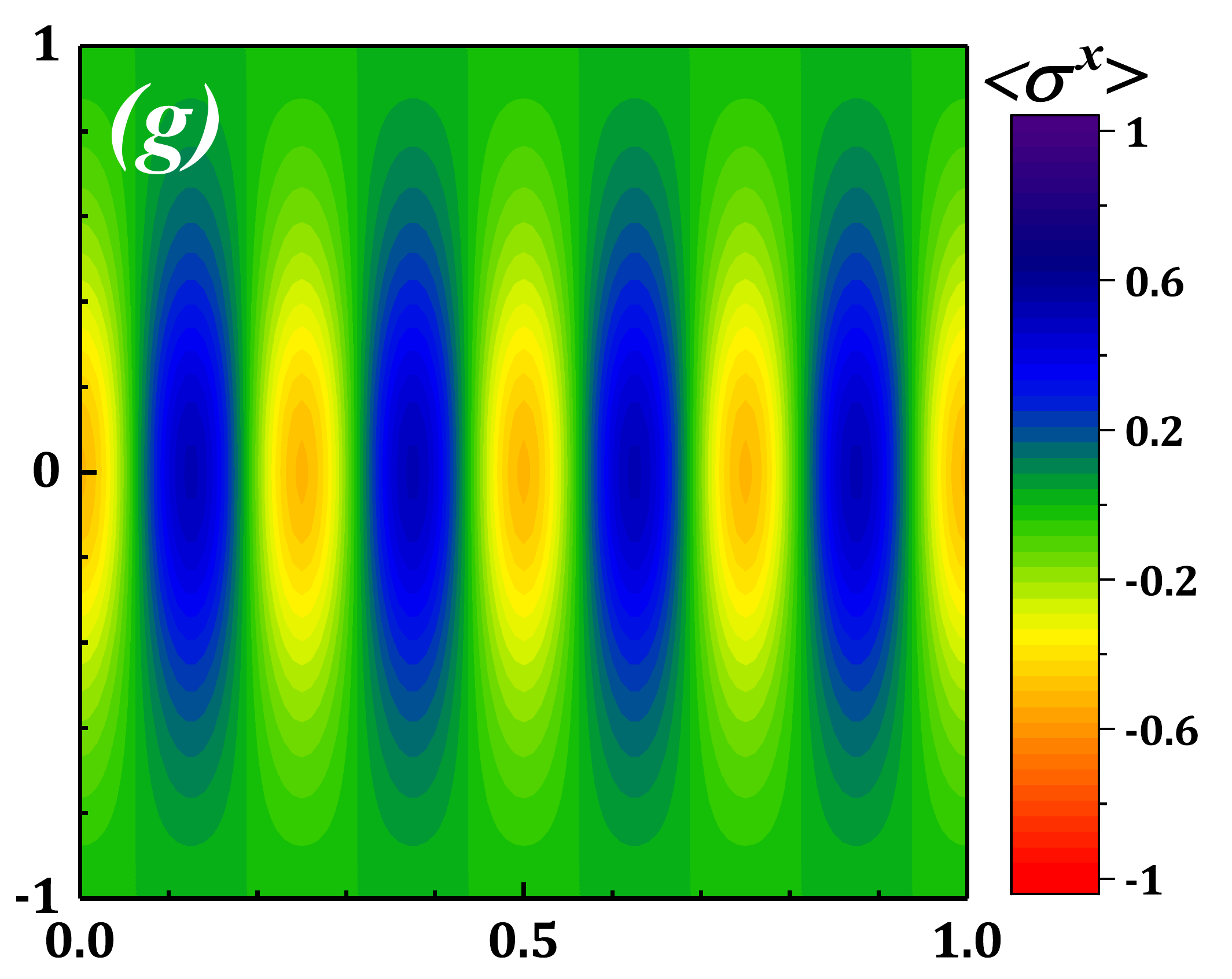}}
\centering
\end{minipage}
\begin{minipage}{\linewidth}
\centerline{\includegraphics[width=0.33\linewidth]{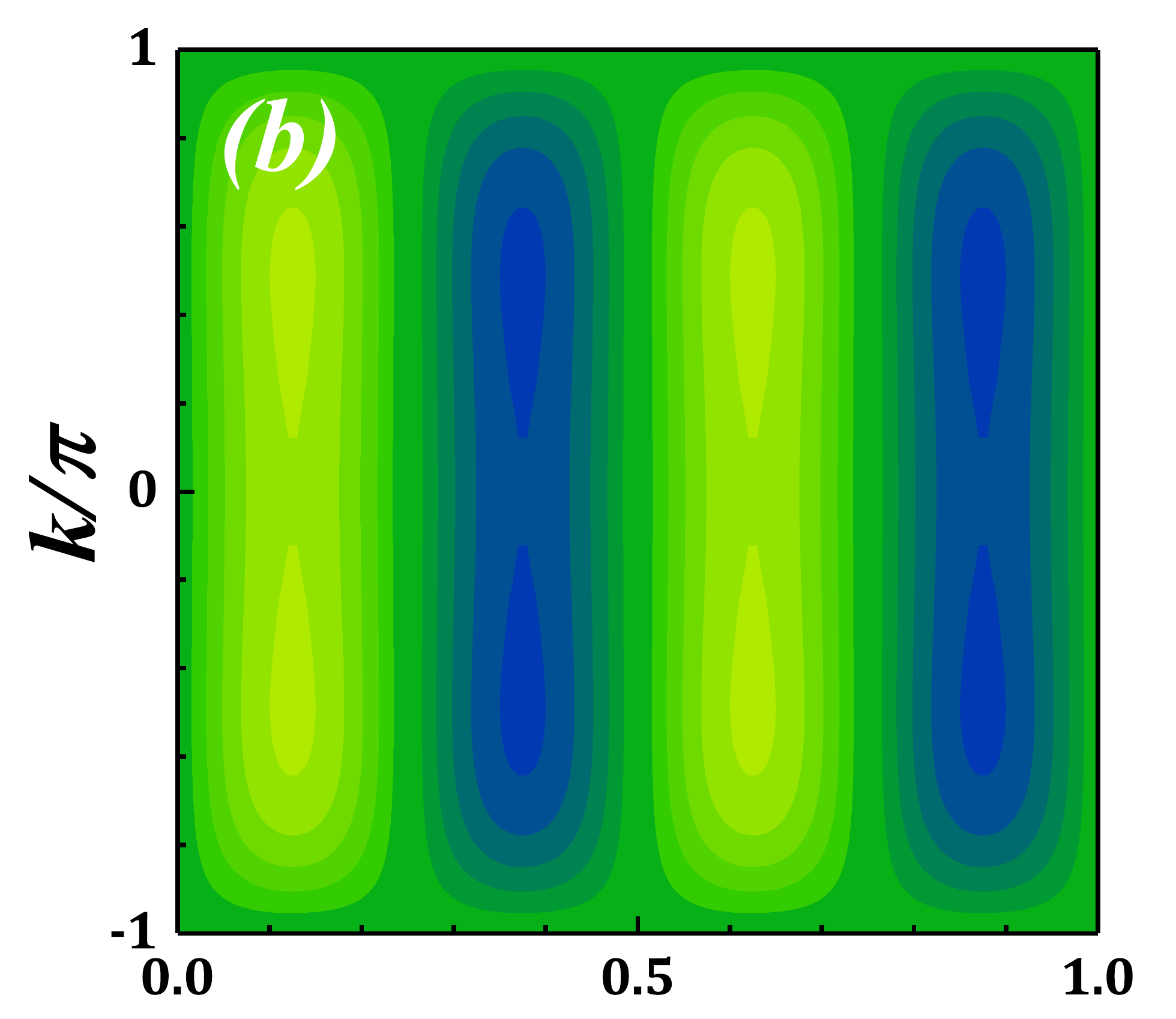}
\includegraphics[width=0.30\linewidth]{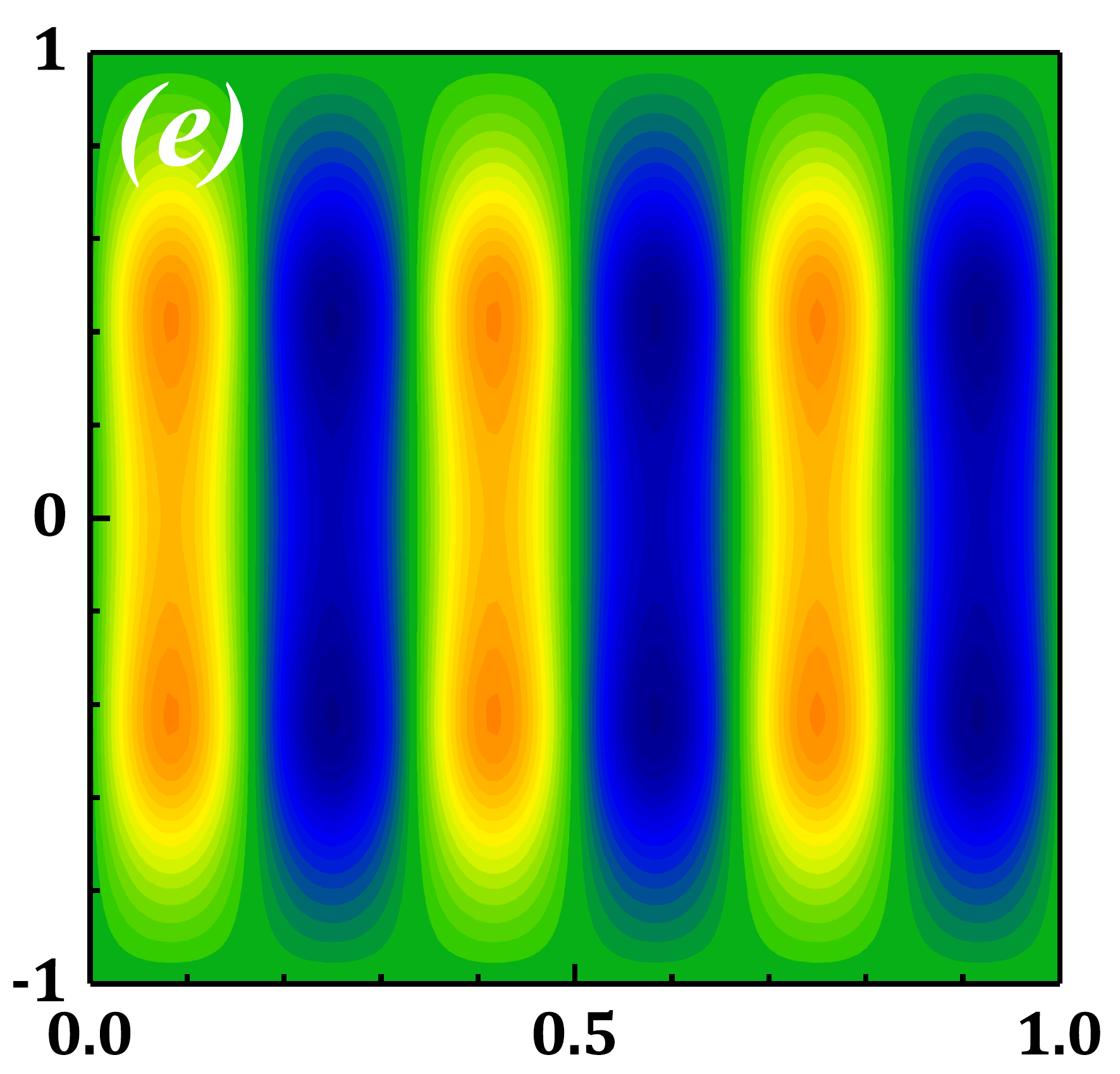}
\includegraphics[width=0.365\linewidth]{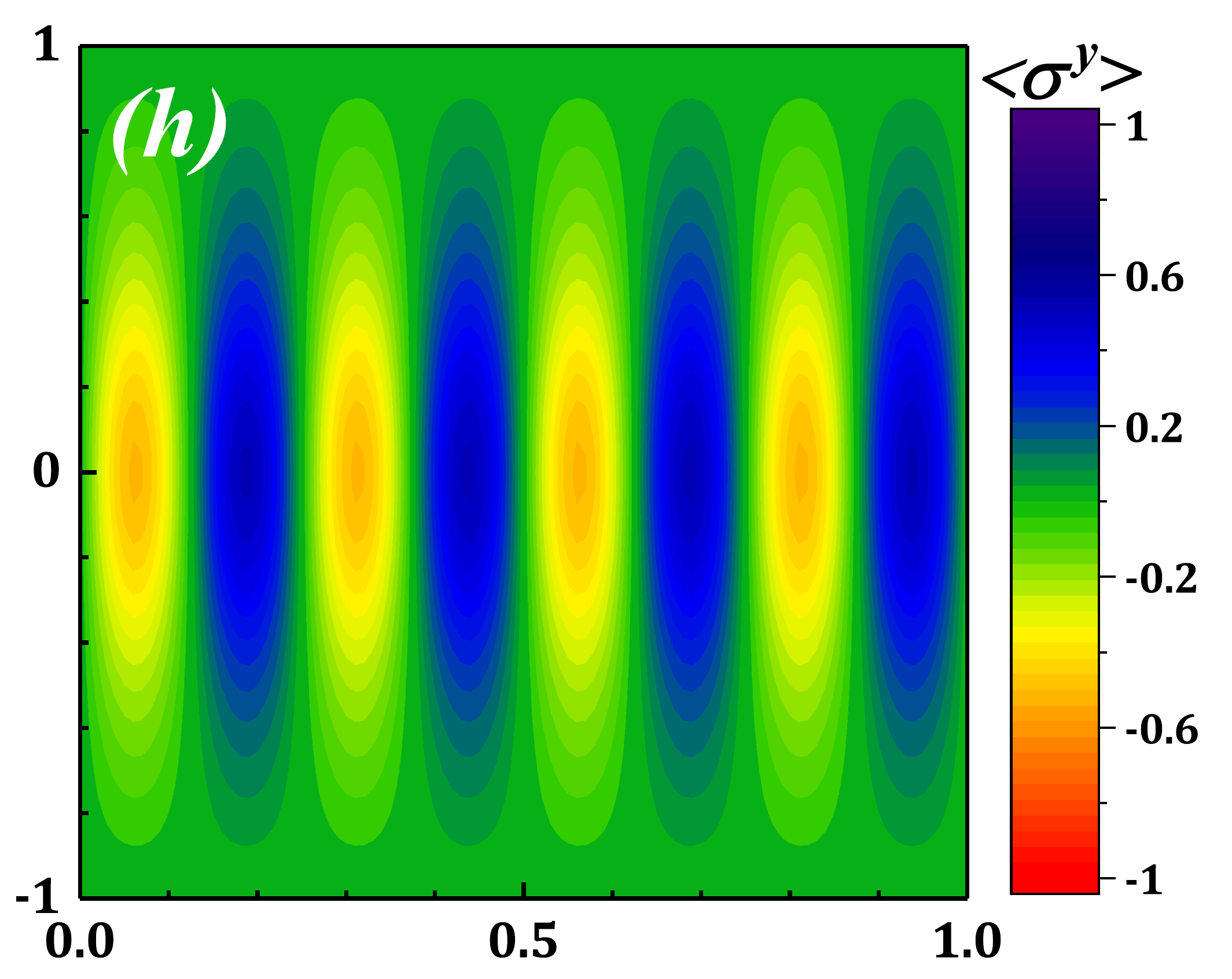}}
\centering
\end{minipage}
\begin{minipage}{\linewidth}
\centerline{\includegraphics[width=0.332\linewidth]{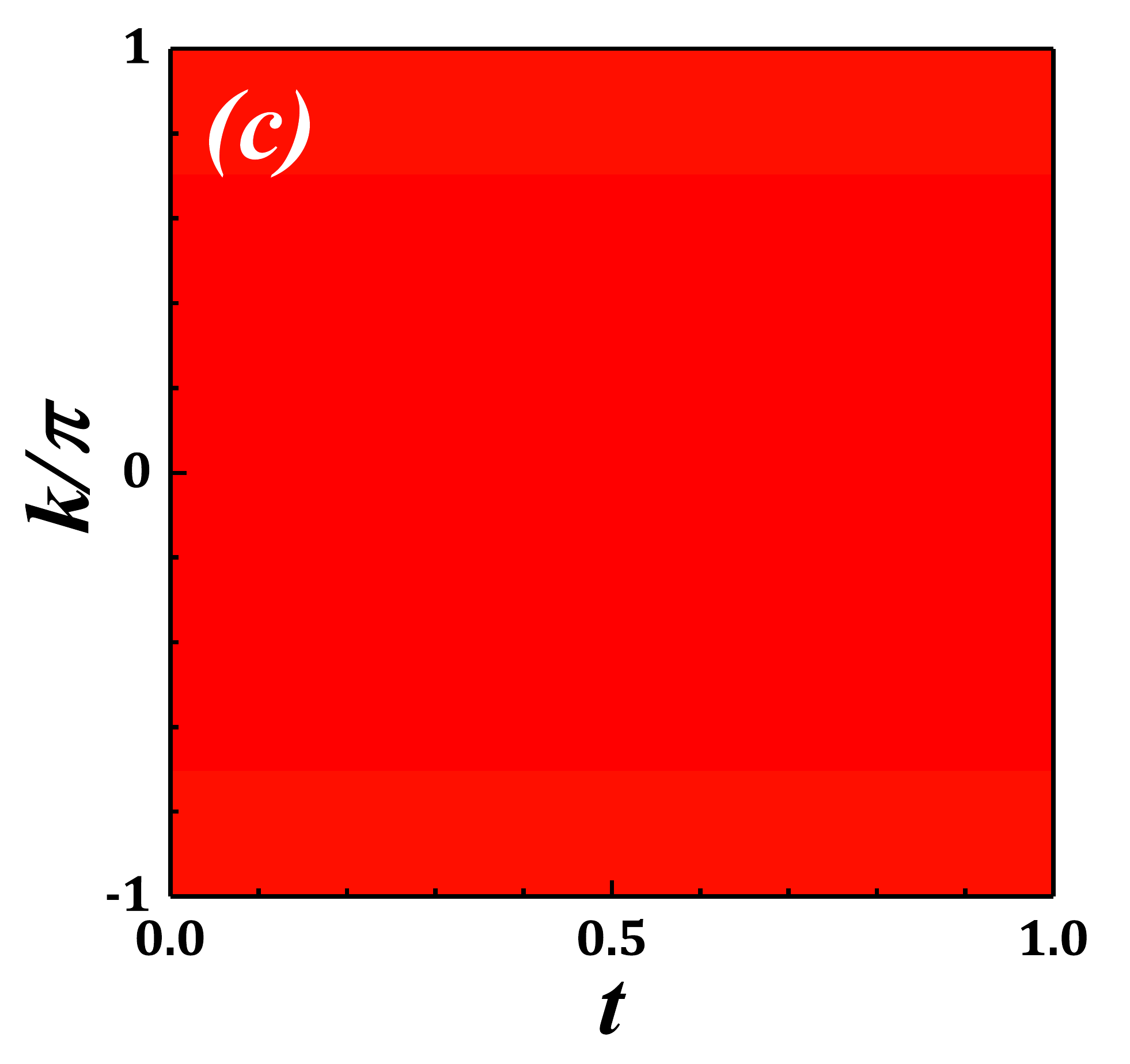}
\includegraphics[width=0.297\linewidth]{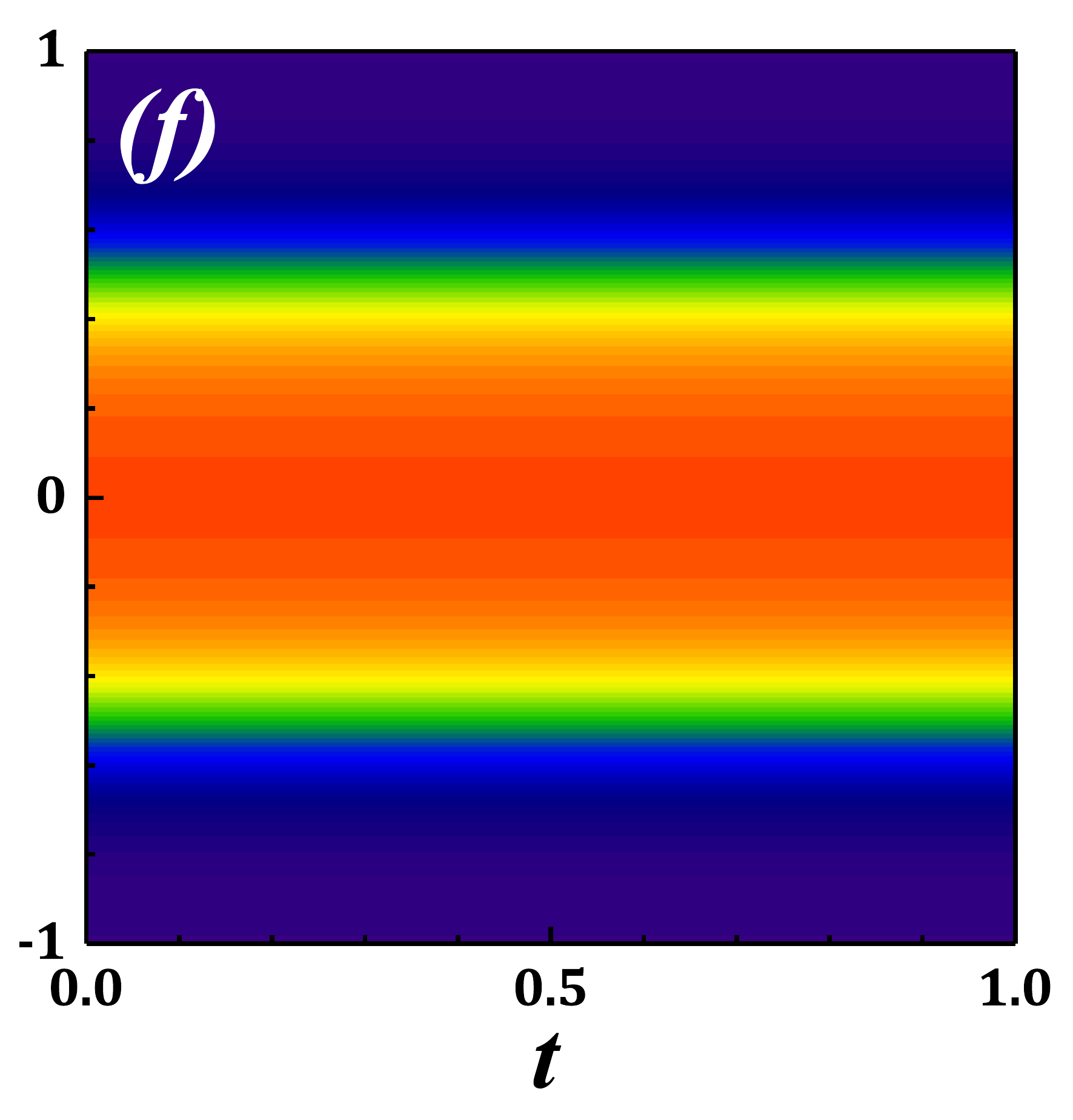}
\includegraphics[width=0.368\linewidth]{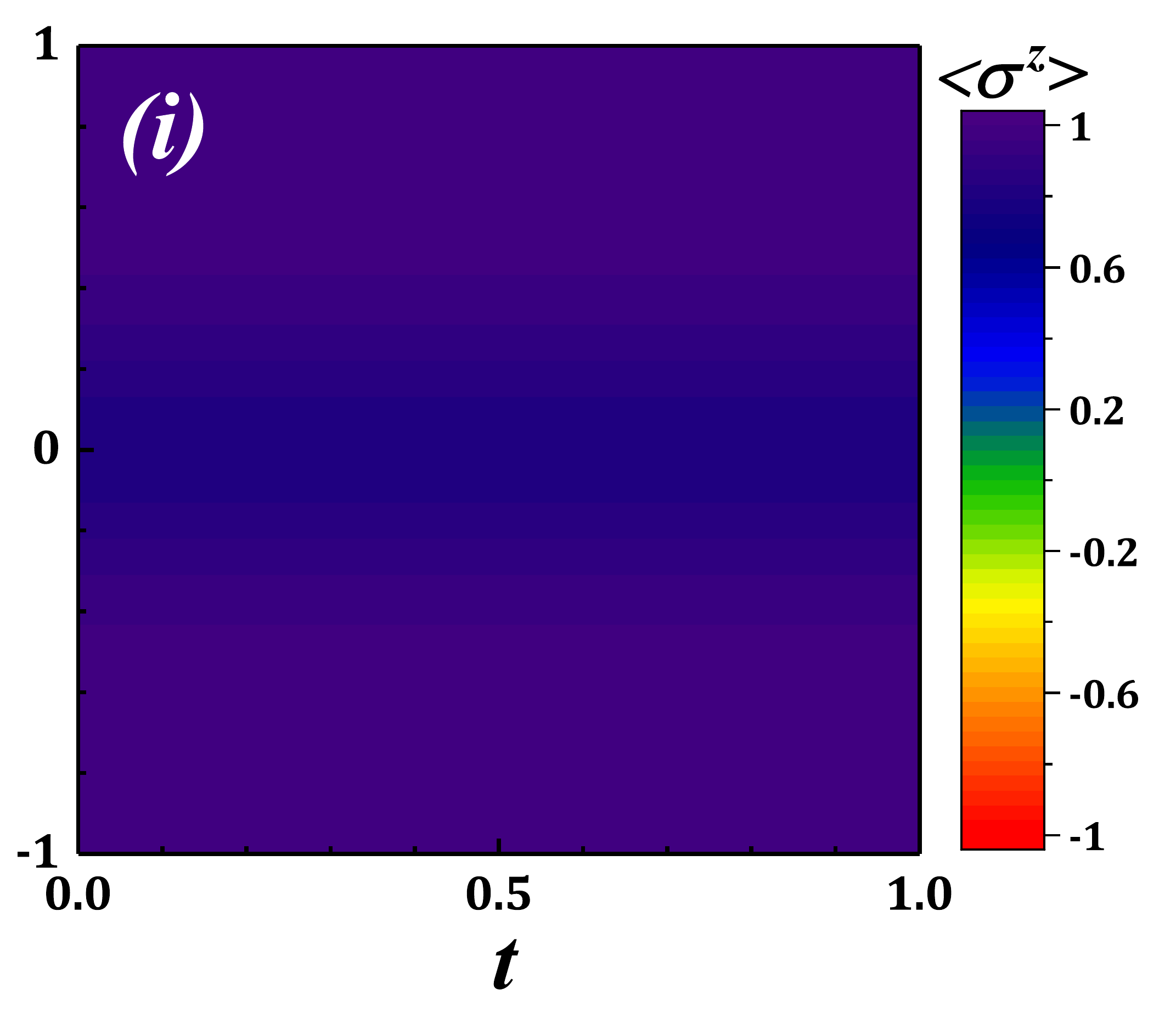}}
\centering
\end{minipage}
\caption{(Color online) Density plot of expectation values of
$\sigma^{x}$, $\sigma^{y}$, and $\sigma^{z}$ for mixed state dynamical phase
transition case versus $t$ and $k$ for $J_{2}=\pi$, $h_{s}=3\pi$, and
(a-c) $\omega=4\pi$, (d-f) $\omega=6\pi$ and (g-i) $\omega=8\pi$.}
\label{fig8}
\end{figure*}
%
However, $A^{\nu}_{k}(\theta)$ is time independent and do not contribute in calculation of Berry curvature. Therefore, the Berry curvature is given by:
%
\begin{eqnarray}
\no
F^{\nu}&=&dA^{\nu}_{k}=\partial_{\theta}A^{\nu}_{k}(t)d\theta\wedge dt
=\frac{\partial A^{\nu}_{k}(t)}{\partial \gamma(\theta)}\frac{\partial \gamma(\theta)}{\partial \theta}d\theta\wedge dt\\
\label{eq37}
&=& \frac{\nu\omega}{2}\frac{\sin(\theta)
-\frac{\omega}{2|\vec{h}_{k}|}\sin(2\theta)}{[1+\frac{\omega^2}{|\vec{h}_{k}|^2}
-\frac{2\omega}{|\vec{h}_{k}|}\cos(\theta)]^{3/2}}d\theta\wedge dt.
\end{eqnarray}
%

Consequently, the Chern number, which is given by the integral of the Berry curvature over the parameter space is in the following form:
%
\begin{eqnarray}
\no
C&=&\frac{1}{2\pi}\int_{0}^{T} dt \int_{0}^{\pi} d\theta F^{-}
=\frac{1}{2\pi}\int_{0}^{T} dt \int_{0}^{\pi} \partial_{\gamma}A^{\nu}_{k}(t)\frac{\partial \gamma(\theta)}{\partial \theta}d\theta\\
\label{eq38}
&=&\Theta (1-\frac{\omega}{\sqrt{h_{xy}^2(k)+h_{z}^2(k)}}).
\end{eqnarray}
%

\section{Expectation values of quasi-spin components in the pure state DQPT\label{A5}}

Time dependent expectation values of quasi-spins $\langle\sigma^{\alpha}\rangle$
in the pure state DQPT are given in the following expressions
%
\begin{eqnarray}
\label{eq39}
&&\langle\psi^{-}_{k}(t)|\sigma^x|\psi^{-}_{k}(t)\rangle= \sin(\gamma_k) \cos(\omega t),\\
\no
&&\langle\psi^{-}_{k}(t)|\sigma^y|\psi^{-}_{k}(t)\rangle= \sin(\gamma_k) \sin(\omega t),\\
\no
&&\langle\psi^{-}_{k}(t)|\sigma^z|\psi^{-}_{k}(t)\rangle= \cos(\gamma_k).
\end{eqnarray}
%
The density plot of the above values have been depicted for several driving frequencies in Fig. \ref{fig7}. As expected,
in the adiabatic regime the expectation values of $\langle\sigma^{\alpha}\rangle$ changes roughly from $-1$ to $1$ (Figs. \ref{fig7}(d)-(f)).
In contrast, for the nonadiabatic regime the expectation values of $\langle\sigma^{\alpha}\rangle$ do not fully cover the interval $[-1,1]$ (Figs. \ref{fig7}(a)-(c) and \ref{fig7}(g)-(i)). We should mention that, the time periodicity of $\langle\sigma^{\alpha}\rangle$  in adiabatic regime coincide with the periodicity of the rate function of LA. Additionally, the expectation values do not decay exponentially in time.

\section{Mixed state dynamical topological order parameter\label{A7}}
To calculate the dynamical topological order parameter in the mixed state, we have to calculate the total phase
and the dynamical phase. The total phase $\phi(k,\beta,t)$ and the dynamical
phase $\phi^{D} (k,\beta,t)$, are given as follows
%
\begin{widetext}
	\begin{eqnarray}
	\no
	&&\phi(k,\beta,t)=Arg\Big[Tr(\rho(k,\beta,0)U(t))\Big]=\tan^{-1}\Big[\frac{\Delta_{k}\cos(\frac{\omega t}{2})\sin(\frac{\Delta_{k}t}{2})+(h_z(k)-\omega)\sin(\frac{\omega t}{2})\cos(\frac{\Delta_{k}t}{2})}
	{\Delta_{k}\cos(\frac{\omega t}{2})\cos(\frac{\Delta_{k}t}{2})-(h_z(k)-\omega)\sin(\frac{\omega t}{2})\sin(\frac{\Delta_{k}t}{2})}\tanh(\beta\Delta_{k}/2)\Big],\\
	\no
	\\
	\label{eq41}
	&&\phi^{D} (k,\beta,t)=-\int_{0}^{t}dt' Tr[\rho(k,\beta,t')H(k,t')]=\frac{\tanh(\beta\Delta_{k}/2) [h_{z}(k)(h_{z}(k)-\omega)+h_{xy}^2(k)]}{2\Delta_{k}}t,
	\end{eqnarray}
\end{widetext}
%
where,
the DTOP for mixed state DPT has been plotted in Fig. \ref{fig6} for different values of $\beta$ and driving frequency $\omega$.

\section{Quasi-spin components in the mixed state\label{A8}}
The time dependent expectation values of quasi-spin components $\langle\sigma^{\alpha}\rangle$
in the mixed state dynamical system are given by the following forms\\
%
{\small
	\begin{eqnarray}
	\no
	&&\langle\sigma^x\rangle= -\frac{\tanh(\beta\Delta_{k}/2)}{\Delta_{k}}h_{xy}(k) \cos(\omega t),\\
	\no
	&&\langle\sigma^y\rangle= -\frac{\tanh(\beta\Delta_{k}/2)}{\Delta_{k}}h_{xy}(k) \sin(\omega t)\\
	\no
	&&\langle\sigma^z\rangle= -\frac{\tanh(\beta\Delta_{k}/2)}{\Delta_{k}}(h_{z}(k)-\omega).
	\end{eqnarray}
}
%
Their corresponding density plots have been illustrated in Fig. \ref{fig8} for several values of driving frequencies at $\beta=1$. As explained before,
in the adiabatic regime the expectation values of $\langle\sigma^{\alpha}\rangle$
sweep the whole range of values, from $-1$ to $1$ (Figs. \ref{fig8}(d)-(f)).
While in the nonadiabatic regime, the expectation values of $\langle\sigma^{\alpha}\rangle$ do not fully cover the interval $[-1,1]$ (Figs. \ref{fig8}(a)-(c) and \ref{fig8}(g)-(i)). It is worth mentioning that, for higher temperatures the variation domain
of expectation values becomes smaller.
Moreover, the expectation values decay exponentially with time at finite temperatures.
We have also observed that the quasi-spin expectation values become approximately constant, when the temperature is comparable to the temperature associated with the energy gap.

\bibliography{References}

\end{document}